\documentclass[fleqn,usenatbib]{mnras}

\usepackage{newtxtext,newtxmath}

\usepackage[T1]{fontenc}

\DeclareRobustCommand{\VAN}[3]{#2}
\let\VANthebibliography\thebibliography
\def\thebibliography{\DeclareRobustCommand{\VAN}[3]{##3}\VANthebibliography}

\usepackage{graphicx}	
\usepackage{amsmath}	

\newcommand{\oiii}{[{\sc O\,iii}]}
\newcommand{\nii}{[{\sc N\,ii}]}
\newcommand{\sii}{[{\sc S\,ii}]}
\newcommand{\hii}{{\sc H\,ii}}

\usepackage{fontawesome}
\usepackage{caption}
\usepackage{subcaption}

\title[Machine Learning Emission-Line Classification]{A Machine Learning Approach to Galactic Emission-Line Region Classification}

\author[C.L. Rhea et al.]{
Carter L. Rhea,$^{1,2}$\thanks{E-mail: carter.rhea@umontrea.ca}
Laurie Rousseau-Nepton,$^{3}$
Ismael Moumen,$^{2,3,4}$
Simon Prunet,$^{5}$ \newauthor
Julie Hlavacek-Larrondo,$^{1}$
Kathryn Grasha,$^{6,7,8}$
Carmelle Robert,$^{4}$
Christophe Morisset,$^{9}$ \newauthor
Grazyna Stasinska,$^{10}$
Natalia Vale-Asari,$^{11,12,13}$
Justine Giroux,$^{4}$
Anna McLeod,$^{14,15}$ \newauthor
Marie-Lou Gendron-Marsolais,$^{16}$
Junfeng Wang,$^{17}$
Joe Lyman,$^{18}$
Laurent Chemin $^{19}$
\\
$^{1}$Département de Physique, Université de Montréal, Succ. Centre-Ville, Montréal, Québec, H3C 3J7, Canada\\
$^{2}$Centre de Recherche en Astrophysique du Québec (CRAQ), Québec, QC, G1V 0A6, Canada\\
$^{3}$Canada-France-Hawaii Telescope, Kamuela, HI, United States\\
$^{4}$Département de physique, de génie physique et d’optique, Université Laval\\
$^{5}$Université Côte d'Azur, Observatoire de la Côte d'Azur, CNRS, Laboratoire Lagrange, France\\
$^{6}$Research School of Astronomy and Astrophysics, Australian National University, Weston Creek, ACT 2611, Australia\\ 
$^{7}$ARC Centre of Excellence for All Sky Astrophysics in 3 Dimensions (ASTRO 3D), Australia\\ 
$^{8}$ ARC DECRA Fellow\\ 
$^{9}$Instituto de Astronomía, Universidad Nacional Autónoma de México, AP 106, 22800 Ensenada, B. C., Mexico\\ 
$^{10}$Laboratoire Univers et Th\'eorie, Observatoire de Paris, Université PSL, Université Paris Cit\'e, CNRS, F-92190 Meudon, France\\
$^{11}$Departamento de F\'{\i}sica--CFM, Universidade Federal de Santa Catarina, C.P.\ 476, 88040-900, Florian\'opolis, SC, Brazil\\ 
$^{12}$School of Physics and Astronomy, University of St Andrews, North Haugh, St Andrews KY16 9SS, UK
$^{13}$Royal Society--Newton Advanced Fellowship\\ 
$^{14}$Centre for Extragalactic Astronomy, Department of Physics, Durham University, South Road,  Durham DH1 3LE, UK\\ 
$^{15}$Institute for Computational Cosmology, Department of Physics, University of Durham, South Road, Durham DH1 3LE, UK\\
$^{16}$European Southern Observatory, Alonso de Co\'ordova 3107, Vitacura, Casilla 19001, Santiago de Chile\\
$^{17}$Department of Astronomy, Xiamen University, Xiamen, Fujian 361005, China
$^{18}$Department of Physics, University of Warwick, Coventry, CV4 7AL, UK\\
$^{19}$Instituto de Astrofisica, Universidad Andres Bello, Fernandez Concha 700, Las Condes, Santiago RM, Chile\\
}

\date{Accepted XXX. Received YYY; in original form ZZZ}

\pubyear{2015}

\begin{document}
\label{firstpage}
\pagerange{\pageref{firstpage}--\pageref{lastpage}}
\maketitle

\begin{abstract}
Diagnostic diagrams of emission-line ratios have been used extensively to categorize extragalactic emission regions; however, these diagnostics are occasionally at odds with each other due to differing definitions.
In this work, we study the applicability of supervised machine-learning techniques to systematically classify emission-line regions from the ratios of certain emission lines. Using the Million Mexican Model database, which contains information from grids of photoionization models using \texttt{cloudy}, and from shock models, we develop training and test sets of emission line fluxes for three key diagnostic ratios. The sets are created for three classifications: classic \hii{} regions, planetary nebulae, and supernova remnants. We train a neural network to classify a region as one of the three classes defined above given three key line ratios that are present both in the SITELLE and MUSE instruments' band-passes: [{\sc O\,iii}]$\lambda5007$/H$\beta$, [{\sc N\,ii}]$\lambda6583$/H$\alpha$, ([{\sc S\,ii}]$\lambda6717$+[{\sc S\,ii}]$\lambda6731$)/H$\alpha$. We also tested the impact of the addition of the [{\sc O\,ii}]$\lambda3726,3729$/[{\sc O\,iii}]$\lambda5007$ line ratio when available for the classification. A maximum luminosity limit is introduced to improve the classification of the planetary nebulae. Furthermore, the network is applied to SITELLE observations of a prominent field of M33. We discuss where the network succeeds and why it fails in certain cases. Our results provide a framework for the use of machine learning as a tool for the classification of extragalactic emission regions. Further work is needed to build more comprehensive training sets and adapt the method to additional observational constraints.
\end{abstract}




\section{Introduction}

Extragalactic emission line regions have been extensively studied over the last hundred years. These studies highlight the different feedback mechanisms responsible for injecting significant amounts of energy in the interstellar medium (ISM) and for ionizing it (i.e. \citealt{veilleux_spectral_1987}; \citealt{osterbrock_astrophysics_1989}). Presently, a vast number of optical observations of extragalactic ionized gas in emission-line regions are available for which we have accurate measurements of the intensity of strong emission lines such as H$\alpha$, H$\beta$, [{\sc N\,ii}]$\lambda$6583, [{\sc S\,ii}]$\lambda6717$, [{\sc S\,ii}]$\lambda6731$, [{\sc O\,iii}]$\lambda5007$, and [{\sc O\,ii}]$\lambda3726,3729$ (e.g. \citealt{baldwin_classification_1981}; \citealt{macalpine_curtis_1981}; \citealt{ivezic_optical_2002}; \citealt{salzer_spectroscopy_2005}; \citealt{kewley_using_2002}; \citealt{moustakas_optical_2010}; \citealt{kewley_understanding_2019}). These regions are generally categorized based on their morphological features (e.g. compact or extended, slope of the luminosity profile, general shape, etc.) or excitation mechanisms (radiative or mechanic). These can be segregated into three main classes for the most common bright emission line regions: {\sc H\,ii} regions photoionized by young hot stars or star clusters (e.g. \citealt{melnick_giant_1987}; \citealt{viallefond_star_1985}), supernova remnants and other shock induced emission regions (e.g. \citealt{danziger_optical_1976}; \citealt{fesen_optical_1985}), and planetary nebulae (e.g. \citealt{oserbrock_planetary_1964}; \citealt{miller_planetary_1974}). 
We do not consider activate galactic nuclei since this work is aimed at spatially resolved observations.

Initially categorized in \citealt{sersic_h_1960} as luminous extragalactic emission, {\sc H\,ii} regions represent an important class of objects (e.g., \citealt{kennicutt_structural_1984}; \citealt{kennicutt_properties_1989}). These regions emerge from giant molecular clouds where a young stellar cluster containing at least one ionizing O or B star is formed through the gravitational collapse of the cloud (e.g., \citealt{osterbrock_astrophysics_1989}). Depending on the properties of the ionizing sources, the region can vary significantly in size, luminosity, and morphology. Recombination lines from hydrogen and helium are predominant in the nebula's optical spectra. Numerous collisionally-excited lines, CELs, emitted by different ions of metals such as oxygen, sulfur, and nitrogen are also present (e.g., \citealt{kewley_optical_2001}; \citealt{kewley_host_2006}; \citealt{baldwin_classification_1981}). 

Meanwhile, planetary nebulae (PNe) are relatively compact objects formed by the gaseous ejecta (stellar envelope) from an evolving low-mass star which is later ionized by the star as it evolves to higher temperatures. Similar to classic \hii{} regions, their optical spectra are dominated by strong recombination and collision lines (e.g., \citealt{oserbrock_planetary_1964}; \citealt{osterbrock_astrophysics_1989}; \citealt{miller_planetary_1974}). 
Supernovae remnants (SNRs) are formed from the gaseous debris scattered following the explosive death of a massive star (e.g., \citealt{fesen_optical_1985}) or thermonuclear runaway in a white dwarf (\citealt{iben_supernovae_1984}). They are ionized both by high-velocity shocks in the ISM and the stellar remnant. Their size, surface brightness, and morphology evolve quickly through time until they blend with the surrounding diffuse ionized gas medium of their host galaxy (e.g.,  \citealt{moumen_3d_2019}; \citealt{smith_optical_1993}; \citealt{woltjer_supernova_1972}). 

The differing ionization mechanisms and underlying physics of these three main classes of emission-line regions manifest themselves distinctly in the relative intensity of the lines of their optical spectra.
Although the original Baldwin-Phillips-Terlevich (BPT; \citealt{baldwin_classification_1981}) diagnostic diagrams have been revised several times, they continue to represent the main emission mechanism characterization tools (e.g., \citealt{veilleux_spectral_1987}; \citealt{kewley_cosmic_2013}). Several prominent theoretical diagnostic lines for these diagrams exist; the most notable are the \cite{kewley_optical_2001} and \cite{kauffmann_host_2003} diagnostics. Using these diagnostic line formulas, observers are able to classify excitation mechanisms. However, discrepancies between the two diagnostic lines, along with issues segregating shocked gas emission from AGN emission, have led to the creation of new classification schemes (\citealt{kewley_host_2006}; \citealt{constantin_clustering_2006}; \citealt{dagostino_new_2019}; \citealt{de_souza_probabilistic_2017}). Moreover, the data used to create the diagnostic lines were determined from integrated spectra of galaxies rather than resolved, parsec-scale objects.

New integral field units (IFUs) are revolutionizing the way in which extragalactic emission-line regions are studied by providing both the spectral and spatial information at the same time (e.g., \citealt{henault_muse_2003}; \citealt{della_bruna_studying_2020}; \citealt{mcleod_impact_2021}; \citealt{kreckel_revised_2017}). Using these instruments, several observatories are deploying large legacy surveys; 
SIGNALS is the Star formation, Ionized Gas, and Nebular Abundances Legacy Survey currently being conducted at the Canada-France-Hawaii Telescope (CFHT; \citealt{rousseau-nepton_signals_2019}). The program uses over 350 hours of observing time on the CFHT's new imaging Fourier Transform Spectrometer, SITELLE (e.g., \citealt{martin_calibrations_2017}; \citealt{baril_commissioning_2016}; \citealt{drissen_sitelle_2019}). SITELLE produces spectral cubes that contain more than 4 million spaxels with a varying \textbf{spectral} resolution (R$\sim$1-10000). This enables a detailed study of extragalactic emission regions in which we can resolve structures spatially and obtain tight constraints on their emission-line ratios.

In this paper, we explore the use of artificial neural networks (ANNs) to categorize extragalactic emission-line objects into {\sc H\,ii} regions, SNRs, and PNe. In $\S$ \ref{sec:meth}, we describe the synthetic data set and the neural network architecture used in our analysis. In $\S$ \ref{sec:results}, we share the results of our network and compare them with traditional classification techniques. We apply the network to a SITELLE field of M33 in $\S$\ref{sec:M33}. In $\S$ \ref{sec:conclusions}, we present the conclusions of our work.
\section{Methodology, Observations, and Simulations}

\subsection{Methodology}
\label{sec:meth} 

In the following section, we outline the databases and methods used to create synthetic emission-line ratios and the machine learning algorithm utilized to classify the ratios into different ionization mechanisms.

\subsection{Synthetic Data}\label{sec:syn}

In order to train and test our classification methodology, a set of emission-line ratios labeled by the region's type (i.e., H{\sc \,ii}, PNe, or SNR) is required. We use the Million Mexican Model Database (3MdB; \citealt{morisset_virtual_2015}; \citealt{alarie_extensive_2019}) and its ancillary databases to construct the training, validation, and test sets). 3MdB contains several grids of simulations that use the photoionization code \texttt{Cloudy} (\texttt{v.17}) to emulate the expected emission from different ionizing sources and their surrounding ISM (\citealt{ferland_2017_2017}). We take the line intensity values of lines of interest (discussed in detail below), which are quoted in \texttt{ergs/s} in the 3MdB. We use the 3MdB project entitled \texttt{BOND} (\citealt{asari_bond_2016}) to obtain lines ratios from classic {\sc H\,ii} regions; similarly, we use the \texttt{PNe} (associated with \citealt{delgado-inglada_ionization_2014}) project to obtain line ratios for planetary nebulae: the data can be found under \texttt{PNe\_2021} in the 3MdB. Since these models were run expressly for this project, we include a brief description here. The primary difference between the previous version, \texttt{PNe\_2020}, and the updated version is that the O/H grid now covers a wider range of values (from -5.46 to -2.96) with finer sampling, and the N/O ratio was left free during the simulations. Together with other minor changes, the new models are more representative of physical planetary nebulae expected to be seen by the SIGNALS collaboration. We further constrain the planetary nebulae sample following the methodology outlined in \cite{delgado-inglada_ionization_2014}.

Regarding the \hii{} regions, we only consider a subset of the entire \texttt{BOND} simulation set in 3MdB since the original set contains models that are not likely to represent \hii{} regions that can be found in nature.
A method to select a subset of the model database, along with a detailed discussion on the necessity to sub-sample the database, is presented for the case of giant \hii{} regions in \citet{amayo_ionization_2021}. In this present study, we created a subset of the model that remains broader, constraining only the physical parameters within the range expected in local galaxies; we retain only the completely filled geometry (i.e., the gas fills the entire volume), which allows us to focus on a younger population of \hii{} regions (the ages are between 1 and 6 Myr; \citealt{asari_bond_2016}; \citealt{cedres_filling_2013}; \citealt{stasinska_excitation_2015}). Moreover, we constrain the ionization parameter, log(U), between $-$3.5 and $-$2.5, the metallicity proxy, 12+log(O/H), between 7.4 and 9.0, and the nitrogen to oxygen ratio, log(N/O), between $-$2 and 0 (\citealt{rousseau-nepton_signals_2019}; \citealt{perez-montero_revisiting_2019}; \citealt{kashino_disentangling_2019}; \citealt{zinchenko_effective_2019}; \citealt{rhea_machine_2020}). This selection ensured that we kept models that could represent the space of parameters for the giant \hii{} regions, faint \hii{} regions, and those lying in uncommon environments. It also includes indirect regions composed of one or a few ionizing stars (O or B stars) since the ionizing spectrum of a stellar population is dominated by the emission of the most massive stars; as the age increases, it ultimately becomes dominated by the late B stars.

\begin{figure}
    \centering
    \includegraphics[width=0.48\textwidth]{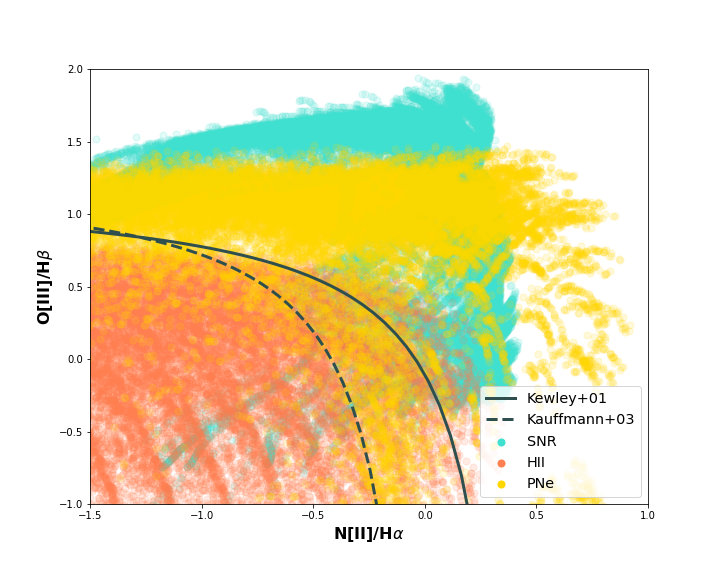}
    \caption{
    \nii{}$\lambda6583$/H$\alpha$ vs. [{\sc O\,iii}]$\lambda5007$/H$\beta$ diagnostic plots. The diagnostic lines from \citet{kewley_optical_2001} and \citet{kauffmann_host_2003} are plotted in black. 
    Values beneath the \citet{kauffmann_host_2003} line are interpreted as photo-ionized while regions above the \citet{kewley_optical_2001} line are considered as shock-ionized regions. Points that lie between the two curves are classified as composite regions.
    The points represent our training, validation, and test sets.
    Figure \ref{fig:BPT1} shows this BPT diagram broken down into each region.
    }
    \label{fig:NIIvsOIII}
\end{figure}

\begin{figure}
    \centering
    \includegraphics[width=0.48\textwidth]{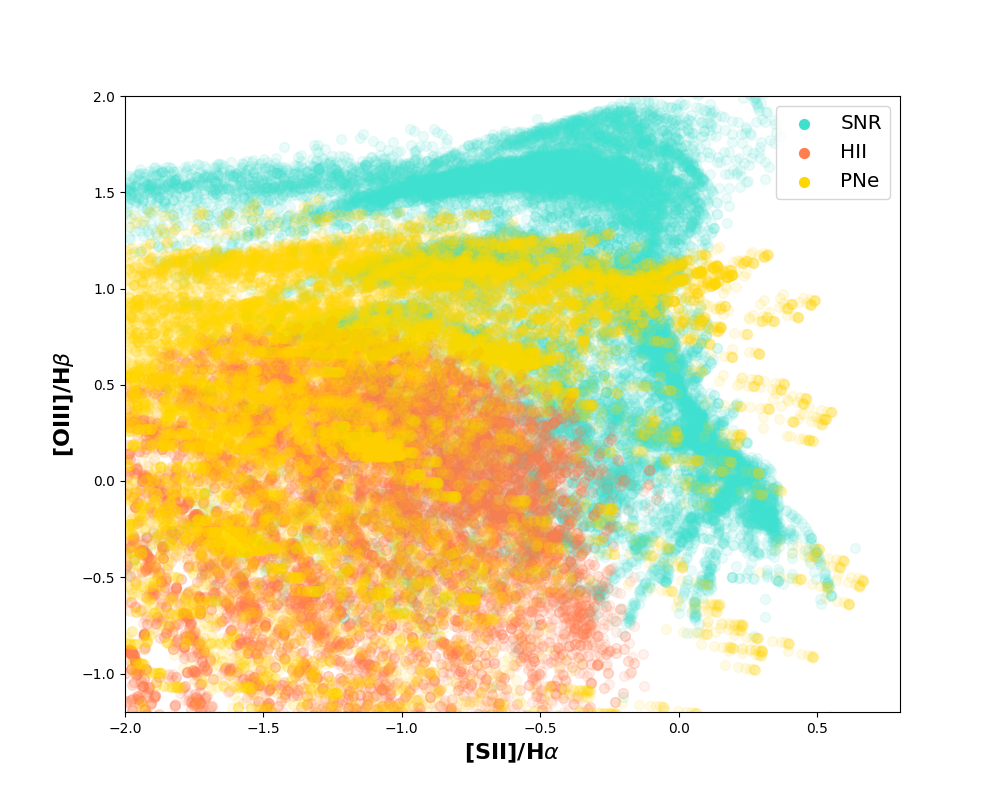}
    \caption{
    \sii{}$\lambda(6731+6716)$/H$\alpha$ vs. \oiii{}$\lambda5007$/H$\beta$ diagnostic plots. The theoretical AGN and LINER/Seyfert 2 lines are plotted in black and are taken from \citet{kewley_host_2006}. The points represent our training, validation, and test sets. Figure \ref{fig:BPT2} shows this BPT diagram broken down into each region.
    }
    \label{fig:SIIvsOIII}
\end{figure}

The supernova remnant emission lines were taken from the 3MdBs\footnote{3MdB shock} table described in \cite{alarie_extensive_2019}.
Although 3MdB is not specifically for supernova remnants, we interpret the simulated values as coming from such objects. We note this may affect subsequent classifications of real data. 
Information pertaining to each grid can be found in the respective project's reference paper. The data of the subgrids are shown in Figures \ref{fig:NIIvsOIII} and \ref{fig:SIIvsOIII} plotted on characteristic BPT diagrams with diagnostic lines overlaid.

Each simulation contains information for thousands of emission lines. This paper focuses on the emission lines available as part of the SIGNALS program. 
Although the classic BPT diagrams require four line ratios ([{\sc O\,iii}]$\lambda5007$/H$\beta$, [{\sc N\,ii}]$\lambda6583$/H$\alpha$, ([{\sc S\,ii}]$\lambda6717$+[{\sc S\,ii}]$\lambda6731$)/H$\alpha$, [{\sc O\,i}]$\lambda6300$/H$\alpha$), we excluded [{\sc O\,i}]$\lambda6300$/H$\alpha$ since [{\sc O\,i}]$\lambda6300$ is outside of the standard SITELLE filters (e.g. \citealt{baldwin_classification_1981}; \citealt{kewley_host_2006}; \citealt{martin_calibrations_2017}). We thus are left with three strong line ratios which are all attainable using the SITELLE filters SN2 (480-520 nm) and SN3 (651-685 nm). 

We retained only the models for which flux of the three strong lines used in classification is higher than 1\% that of the H$\alpha$ emission. This ensures that all lines are visible in the original emission spectra and can be detected in the SIGNALS program (e.g. \citealt{rousseau-nepton_signals_2019}).
There are 31,911 {\sc H\,ii} regions, 266,238 PNe, and 355,683 SNR. We randomly sampled 30,000 models from each grid to reduce training time and have consistently sized samples. To avoid a bias in the network towards data with more training set data, we have set the number of PNe and SNR samples to be the same order of magnitude as the \hii{} sample; Our testing revealed that 30,000 samples is sufficient to well-classify both the PNe and SNR regions without artificially injecting bias. 
The reduction in sample size does not affect results since the grid sampling remains mostly uniform over the target parameter space. 
We further tested the effect of random sampling by rerunning our analysis ten times with ten differently sampled datasets.

\subsection{Artificial Neural Network} \label{sec:ann}

Artificial neural networks, and their closely related counterpart, convolutional neural networks, are becoming ubiquitous in astronomical applications due to their versatility and speed (e.g., \citealt{bertin_classification_1994} ;\citealt{baron_machine_2019}; \citealt{uzeirbegovic_eigengalaxies_2020}; \citealt{shatskiy_neural_2019}; \citealt{biswas_classification_2018}). In this paper, we explore the use of an artificial neural network to classify extragalactic emission regions into {\sc H\,ii} regions, PNe, and SNRs. The ANN takes the [{\sc O\,iii}]$\lambda5007$/H$\beta$, [{\sc N\,ii}]$\lambda6583$/H$\alpha$, ([{\sc S\,ii}]$\lambda6717$+[{\sc S\,ii}]$\lambda6731$)/H$\alpha$, 
ratios as inputs and outputs the most likely categorization and its corresponding probability. Following standard methods, we use 70\% of the synthetic data for the training set, 20\% for the validation set, and 10\% for the test set (e.g., \citealt{breiman_random_2001}).

The neural network was built using \texttt{tensorflow (v2.4.0)} (\citealt{abadi_tensorflow_2015}) implemented in \texttt{python (v3.6.0)}.
A standard grid-search algorithm to determine the number of layers and nodes within each layer, implemented in \texttt{sklearn (v0.23)}, revealed three layers with 126, 256, and 128 nodes, respectively, to be optimal. In each layer, the nodes are subject to the standard \texttt{relu} activation function; we use the categorical cross-entropy loss function. We implement two dropout layers of 25\% in between the first and second layers and the second and third layers. Since the \texttt{softmax} activation function is used on the last layer (i.e., the output layer), each node of the output layer has a final logit value associated with the probability of the classification being an \hii{} region (0), a planetary nebula (1), or a supernova remnant (2).
We also employ the Adam optimizer algorithm described in \cite{kingma_adam_2017}.  

Further network details can be seen in the demo code at \url{https://github.com/sitelle-signals/Pamplemousse}. 

\begin{figure*}[!h]
    \centering
    \subfloat{\includegraphics[width=0.45\textwidth]{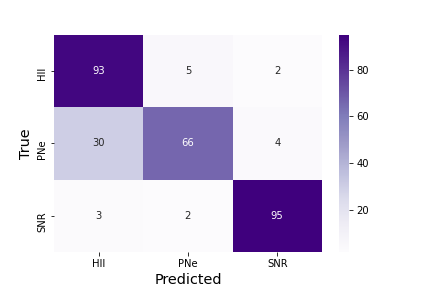}}
    \qquad
    \subfloat{\includegraphics[width=0.45\textwidth]{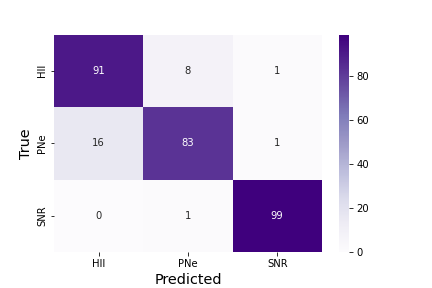}}
    \caption{
    Confusion matrices for the artificial neural networks applied to the test set using three (left) and four (right) line ratios as input parameters to distinguish between \hii{} regions, supernova remnants, and planetary nebulae. The left-hand side confusion matrix is based on the neural network trained on [{\sc O\,iii}]$\lambda5007$/H$\beta$, [{\sc N\,ii}]$\lambda6583$/H$\alpha$, ([{\sc S\,ii}]$\lambda6717$+[{\sc S\,ii}]$\lambda6731$)/H$\alpha$ while the right contains these three ratios and additionally the [{\sc O\,ii}]$\lambda3726,3729$/[{\sc O\,iii}]$\lambda5007$ line ratio.
     }
    \label{fig:ConfusionMatrices}
\end{figure*}

\section{Results and Discussion}\label{sec:results}

\subsection{Confusion Matrix}\label{sec:results-cm}

Figure \ref{fig:ConfusionMatrices} (left) visualizes the ability of the network to accurately categorize emission regions into either classic \hii{} regions, planetary nebulae, or supernova remnants using the three-line ratios ([{\sc O\,iii}]$\lambda5007$/H$\beta$, [{\sc N\,ii}]$\lambda6583$/H$\alpha$, ([{\sc S\,ii}]$\lambda6717$+[{\sc S\,ii}]$\lambda6731$)/H$\alpha$) as inputs to the network.
We stress that the data used for these line ratios (and thus the confusion matrix results) are valid only for data well-sampled by the test set. If, in reality, the test set is not representative of observation, then the results are expected to be worse than what is reported here.
Each element of the confusion matrix can be interpreted as the efficiency (percentage of accurate classifications) at which the neural network categorizes the emission region \textit{y} to their true type \textit{x}. For example, the first row indicates that the network correctly categorizes \hii{} regions 93\% of the time but that it misclassifies \hii{} as either PNe or SNR 5\% and 2\% of the time, respectively. Similarly, PNe regions are correctly categorized 66\% of the time and are misclassified as either \hii{} regions or supernova remnants 30\% and 4\% of the time, respectively.
Finally, supernova remnants are correctly classified 95\% of the time and are misclassified as \hii{} regions and planetary nebulae 3\% and 2\% of the time, respectively. 
A diagonal confusion matrix, such as we have, indicates that the network is correctly classifying the emission regions the overwhelming majority of the time. 

Additionally, we report the confusion matrix (figure \ref{fig:ConfusionMatrices} right) when we incorporate a fourth line ratio, [{\sc O\,ii}]$\lambda3726,3729$/[{\sc O\,iii}]$\lambda5007$, into the input vector of the network. As evidenced by the figure, the classification accuracy for planetary remnants increases significantly (from 66\% to 83\%), while the classification accuracy for the other two region types remains consistent.

However, we draw attention to the high level of planetary nebulae misclassified as \hii{} regions; this incorrect categorization is likely due to the confusion in this restricted three parameters input space of line ratios between \hii{} regions and planetary nebulae. A possible avenue to solve this entanglement is integrating a luminosity threshold for the PNe. This will be discussed further in \S \ref{Inclusion}.

\subsection{Portability to Other Instruments}
Although the methodology described in this paper has been applied only to line ratios calculated from SITELLE data cubes, the algorithm is instrument-agnostic. Since the method requires three-line ratios, the only requirement is that the instrument captures a signal from $4861$\AA \ to $6731$\AA. Since several instruments do not capture the [{\sc O\,ii}]$\lambda3726/3729$ lines (such as the  Multi Unit Spectroscopic Explorer, MUSE, instrument; \citealt{bacon_muse_2010}), we opted not to use them as inputs for our classifier. Nonetheless, we have demonstrated that the network obtains better classification accuracies using the [{\sc O\,ii}]$\lambda3726,3729$/[{\sc O\,iii}]$\lambda5007$ line ratios. With this in mind, adding additional-relevant line ratios and/or other observational constraints (such as the physical relationship between luminosity and line ratios, size and luminosity, reducing when known the range of metallicity of the training set, etc.) could further increase the performance of the network.

\subsection{Inclusion of Information in the Classification Input Vector}
\label{Inclusion}
\begin{figure}
    \centering
    \includegraphics[width=0.48\textwidth]{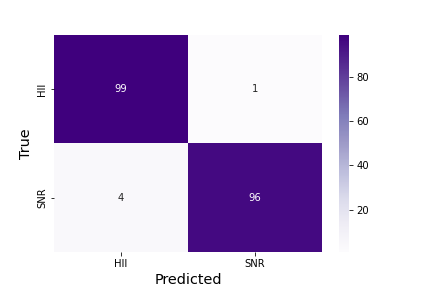}
    \caption{Confusion Matrix for the artificial neural network applied to the test set using the three line ratios ([{\sc O\,iii}]$\lambda5007$/H$\beta$, [{\sc N\,ii}]$\lambda6583$/H$\alpha$, ([{\sc S\,ii}]$\lambda6717$+[{\sc S\,ii}]$\lambda6731$)/H$\alpha$) as input parameters to distinguish between only \hii{} regions and supernova remnants. The values quoted represent the classification percentage.}
    \label{fig:confusion_2}
\end{figure}

In this work, we only considered the three or four-line ratios present in standard BPT plots for our input vector; however, we note that the efficacy of the classifications may be improved by including other line ratios or equivalent widths. Since this work focuses on improving the existing BPT structure, we chose not to include other line ratios. We note, though, that work is currently being done to use unsupervised machine learning algorithms to explore other line ratios (Moumen et al.\ in prep.). 
We note that H$\alpha$ luminosity can be used as a prior to putting additional constrain on the planetary nebulae. Although, at the moment, this is not possible with the current parameters included in 3MdB, we propose the following avenue: incorporate a luminosity threshold above which a region could not be a planetary nebula. As stated by \cite{delgado-inglada_study_2020}, the dust-corrected H$\alpha$ (or H$\beta$) maximum luminosity can be used as a relevant threshold since it does not depend on the PNe metallicities. We define the H$\alpha$ log-luminosity limit for the PNe at 36.0$\pm$ 0.1, the value derived by \cite{delgado-inglada_study_2020} using a sample of 500 extragalactic PNe \textbf{covering} various environments. It is also consistent with values reported by others in the literature, including \cite{martin_sitelle_2018} and \cite{braun_physical_1992}.  
Any region with a H$\alpha$ luminosity above this limit cannot be a planetary nebula. \textbf{If the region is below this limit, we use the classifier that distinguishes between planetary nebulae, supernova remnants, and \hii{} regions.}
Otherwise, we use a classifier trained only to distinguish SNRs and \hii{} regions. Therefore, we train an additional network using the same architecture described above to distinguish only \hii{} regions and SNR.
Figure \ref{fig:confusion_2} demonstrates that the network excels as distinguishing between SNRs and \hii{} regions.

Similar cuts could be applied for supernova remnants (e.g., low luminosity cut in \citealt{delgado-inglada_study_2020}) or high luminosity cut using the brightest SNR observed in the local universe. We do not believe a low-luminosity cut should be used since SNR fade in intensity with time until their luminosity passes the detection limit of the observational configuration. On the other hand, a maximum luminosity could be considered, as it is for the PNe. From the literature, different values can be found: 
\textbf{4.96$\times 10^{37}$ erg/s} in \citealt{winkler_vizier_2017}. Nevertheless, these values are observational and not theoretical; One could expect an even higher maximum luminosity for supernovae surrounded by a denser ISM. 
Since SNRs are not a point-like source compared to extragalactic unresolved PNe, their emission is often embedded in \hii{} regions that can be significantly brighter. In these cases, their estimated luminosity is bias towards higher values. Therefore, we decided not to use a maximum luminosity threshold for the SNR in this work. This decision is based on the possibility of evaluating SNR contamination in \hii{} regions using the method.
We will see later that three of the observed and confirmed SNRs used to test our method are above these observational thresholds and also coincide with objects embedded in \hii{} regions. Additional elements regarding this aspect will be present in the \S \ref{results}. It is also important to mention, as investigated in \cite{vale_asari_role_2021}, that line widths could also be included to help further distinguish regions. The addition of these supplementary input parameters is the subject of future work.

\subsection{Addition of Noise in the Training Set to Fill in Modeling Gaps}\label{sec:noise}
Although the models obtained from 3MdB are extensive and include various scenarios, they are built on grids of model parameters. Therefore, gaps may exist in the distribution of parameters; these gaps can be seen in Figures \ref{fig:SIIvsOIII} and \ref{fig:NIIvsOIII}. In order to explore the effect of these gaps on our results, we add random noise to the synthetic data. In doing so, we draw the line ratio values randomly from a Gaussian distribution centered on the line ratio value from 3MdB with sigma values of 1\%, 2\%, and 5\% the line ratio value. Upon retraining the algorithm for each level of noise, we obtain similar accuracy values for each category. 


\subsection{Application to Diffuse Ionized Gas}
\begin{figure}
    \centering
    \includegraphics[width=0.49\textwidth]{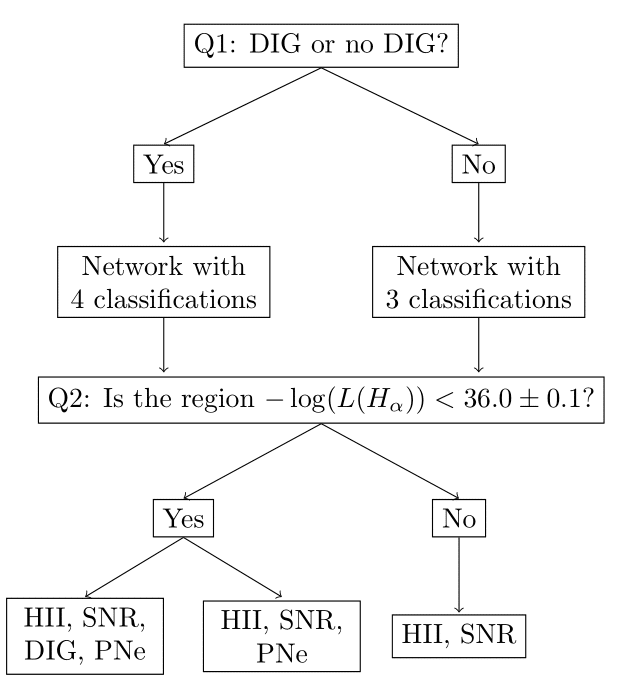}
    \caption{Tree diagram representing the possible classification routes based on the region's luminosity and the chosen classification scheme.}
    \label{fig:BPTDecision}
\end{figure}

\begin{figure}
    \centering
    \includegraphics[width=0.49\textwidth]{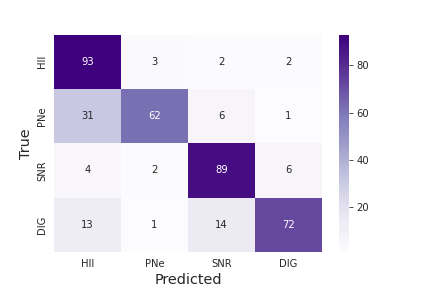}
    \caption{Confusion matrix including DIG as a fourth option using the three standard line ratios as input parameters.}
    \label{fig:DIG}
\end{figure}

In addition to classical \hii{} regions, supernova remnants, and planetary nebula, diffuse ionized gas (DIG) is an important feature of the ISM in many galaxies (e.g., \citealt{reynolds_measurement_1984}; \citealt{walterbos_diffuse_1994}; \citealt{haffner_warm_2009}). Therefore, we constructed a fourth classification set based on DIG simulations from 3MdB using the same methodology discussed in $\S$\ref{sec:syn}. We use the \texttt{DIG\_HR} database and filter out \hii{} regions by only retaining regions for which \texttt{phi\_OB} $<$ 4.5\footnote{\texttt{phi\_OB} is the surface flux of the OB star}. We retrain the artificial neural network developed in $\S$\ref{sec:ann} with DIG regions as a fourth classification. We provide the confusion matrix in figure \ref{fig:DIG}; the figure reveals that the network retains its accuracy of determining \hii{} regions, planetary nebulae, and supernova remnants while achieving an accuracy of over 70\% in classifying DIGs. Incorrectly classified DIGs are equally split between supernova remnant and \hii{} region classifications. While this level of accuracy is acceptable for many cases, we do not use this network in future sections. The authors also note that extensive work has been done to successfully distinguish DIG regions from \hii{} regions using the equivalent width of the H$\alpha$ emission (e.g., \citealt{lacerda_diffuse_2018}). Additionally, we supply users with a decision tree (see figure \ref{fig:BPTDecision}), which can be applied to determine which network to use.

\section{Application to M33}\label{sec:M33}

In order to test the method with observations, we use a field from the SITELLE instrument on the galaxy M33.
These observations were conducted for the SIGNALS program and are ideal for this present study since previous identification of SNR and PNe sources abounds in the literature and the high spatial resolution enables a precise selection of the emission area of the candidates.

\subsection{The Data}
\begin{figure*}
    \centering
    \includegraphics[width=0.9\textwidth]{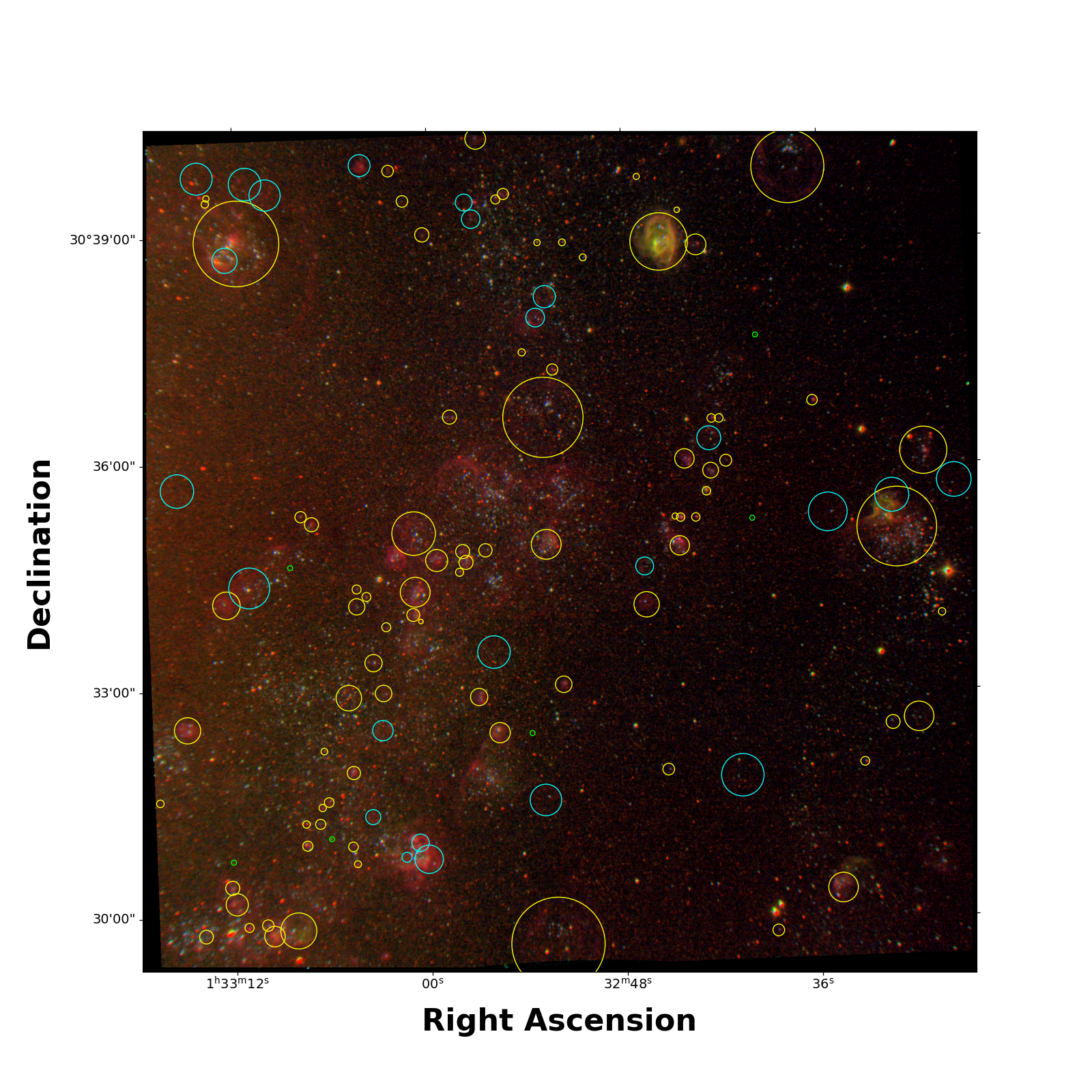}
    \caption{SITELLE Field 7 of M33. This image is a combination of the SN1 deep frame (blue), the SN2 deep frame (green), and the SN3 deep frame (red).
    Here, we can see the location and size (corresponding to the circle size) of HII regions (yellow), planetary nebulae (green), and supernova remnants (cyan). This figure was made by using the \texttt{make\_lupton\_rgb} function implemented in \texttt{astropy} (\citealt{make_lupton_rgb}; \citealt{robitaille_astropy_2013}).
    }
    \label{fig:M33}
\end{figure*}

We extracted the PNe and SNR sources from the \cite{ciardullo_planetary_2004} and \cite{long_mmt_2018} catalogs, respectively. In order to also include a significant fraction of the \hii{} regions visible in the field, we used a combination of the information provided from an emission map produced by adding all emission line maps together and the known position of the PNe and SNR.  Using the ds9 tools, we handpicked a large fraction of the \hii{} regions while keeping only the good candidates, excluding regions with not well-defined contours or obvious contamination. Figure \ref{fig:M33} shows all regions selected in the final test set. Some \hii{} regions harbor a small overlap with a previously identified SNR. We tried to minimize the impact of these overlaps and avoid selecting \hii{} regions that were particularly well blended with the SNRs. It is also important to note that the SNR radii were selected using the visible morphology of the SNR optical component (expanding shell). Nevertheless, we did not exclude known SNRs that are embedded in \hii{} regions emission. We will discuss later in this section the impact of this inherent contamination.

Our literature review revealed 84 \hii{} regions, 6 planetary nebulae, and 24 supernova remnants (see figure \ref{fig:M33}). However, upon further investigation of the spectra captured by SITELLE of these regions, we rejected 8 \hii{} regions, 4 planetary nebulae, and 2 supernova remnants as having insufficient signal-to-noise\footnote{A signal-to-noise as calculated by \texttt{LUCI} below 2 is considered insufficient.} to properly fit the emission lines.
In order to extract the line ratios for these regions, we used the spectral analysis software \texttt{LUCI}. Using SN1 ($R\sim1800$), SN2 ($R\sim1800$), and SN3 ($R\sim5000$) observations of the M33 field (PI: Laurie Rousseau-Nepton), we fit the following strong emission-lines using a \texttt{sincgauss} function: H$\alpha$, H$\beta$, [{\sc N\,ii}]$\lambda$6583, [{\sc S\,ii}]$\lambda6717$, [{\sc S\,ii}]$\lambda6731$, [{\sc O\,iii}]$\lambda5007$, [{\sc O\,iii}]$\lambda4959$, [{\sc O\,ii}]$\lambda3726$, and [{\sc O\,ii}]$\lambda3729$. After calculating the flux of each line, we applied deredenning by calculating the Balmer decrement and using it in conjunction with a deredenning law.
\begin{equation}
     F_{0,\lambda} = F_{\text{obs}, \lambda} \, e^{\tau_\lambda} = F_{\text{obs}, \lambda} \, e^{\tau_V \, q_\lambda},
\end{equation}
where $F_{\text{obs}, \lambda}$ is the observed flux, $\tau_\lambda$ is the optical depth at a given wavelength, $\tau_V$ is the optical depth in the $V$-band, and the shape of the dust attenuation curve is parametrized by $q_\lambda \equiv \tau_\lambda/\tau_V$.

We use the Cardelli, Clayton \& Mathis (1989) attenuation law with $R_V=3.1$.
We use the Balmer decrement, defined as $B_d=F_\mathrm{obs,H\alpha}/F_\mathrm{obs,H\beta}$, to calculate $\tau_V$:
\begin{equation}
    \tau_V = \frac{1}{q_{H\beta}-q_{H\alpha}}\ln{\frac{B_d}{B_{d,in}}},
\end{equation}
where $B_{d,in}$ is the intrinsic Balmer decrement which is assumed to be 2.87.

\begin{figure}
    \centering
    \includegraphics[width=0.5\textwidth]{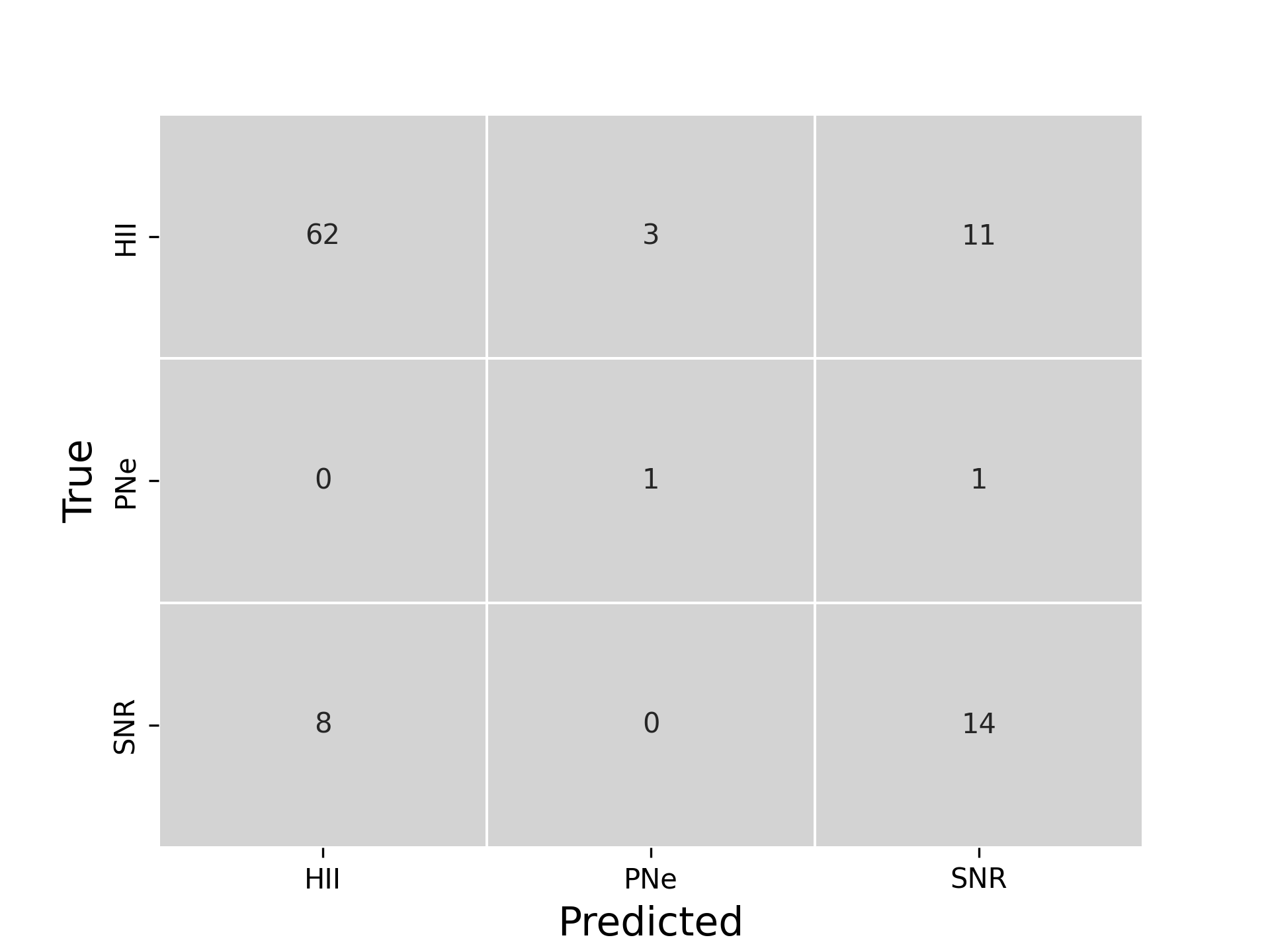}
    \caption{Confusion matrix for M33 Field 7 using a neural network trained with three input parameters: [{\sc O\,iii}]$\lambda5007$/H$\beta$, [{\sc N\,ii}]$\lambda6583$/H$\alpha$, ([{\sc S\,ii}]$\lambda6717$+[{\sc S\,ii}]$\lambda6731$)/H$\alpha$. Values quoted are the number of regions. We only retained regions for which the spectra in all three filters were adequate.}
    \label{fig:M33-field7-confusion}
\end{figure}

\begin{figure*}[!h]
    \centering
    \includegraphics[width=1.0\textwidth]{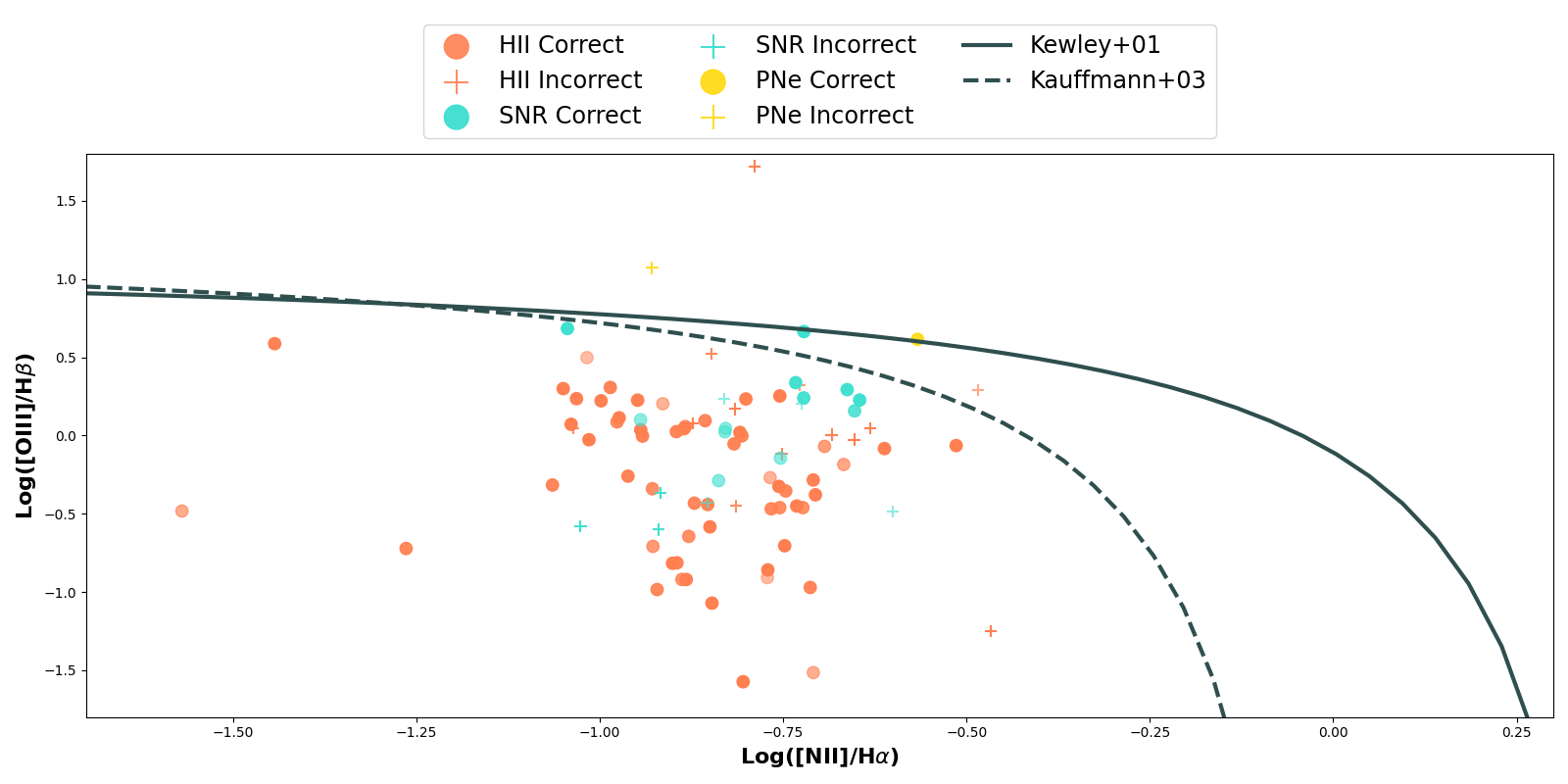}
    \caption{
    The BPT diagram of log([{\sc N\,ii}]/H$\alpha$) vs. log([{\sc O\,iii}]/H$\beta$)  for the regions identified in a SITELLE field of M33. The dashed and solid lines represent the standard \citet{kewley_optical_2001} and \citet{kauffmann_host_2003} delineations. The emission region type is designated by differing colors. Additionally, the correct and incorrect classification (True and False in the legend) are indicated by circles and crosses, respectively. The points' opacity corresponds to the probability of the classification. The lowest probability value is approximately 60\%. We stress that this plot is not to be taken as a comparison between the standard BPT diagnostic cuts and our methodology.
    }
    \label{fig:M33-field7-kk}
\end{figure*}

\begin{figure*}[!h]
    \centering
    \includegraphics[width=1.0\textwidth]{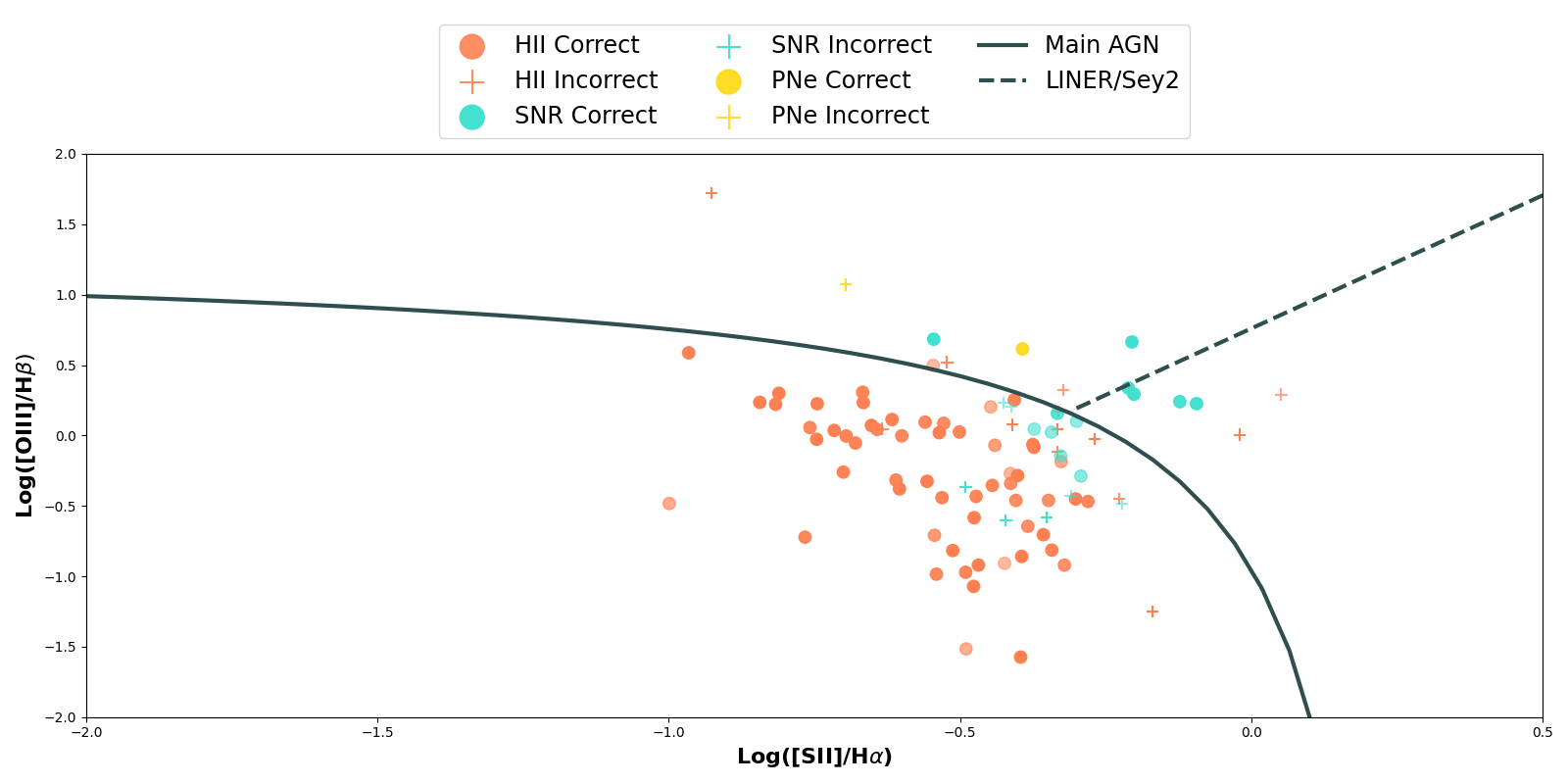}
    \caption{The BPT diagram of log([{\sc S\,ii}]/H$\alpha$) vs. log([{\sc O\,iii}]/H$\beta$)  for the regions identified in a SITELLE field of M33. The dashed and solid lines represent the standard AGN and LINER delineations. The emission region type is designated by differing colors. Additionally, the correct and incorrect classification (True and False in the legend) are indicated by circles and crosses, respectively. The points' opacity corresponds to the probability of the classification. The lowest probability value is approximately 60\%. We stress that this plot is not to be taken as a comparison between the standard BPT diagnostic cuts and our methodology.}
    \label{fig:M33-field7-al}
\end{figure*}

\begin{figure*}[!h]
    \centering
    \includegraphics[width=1.0\textwidth]{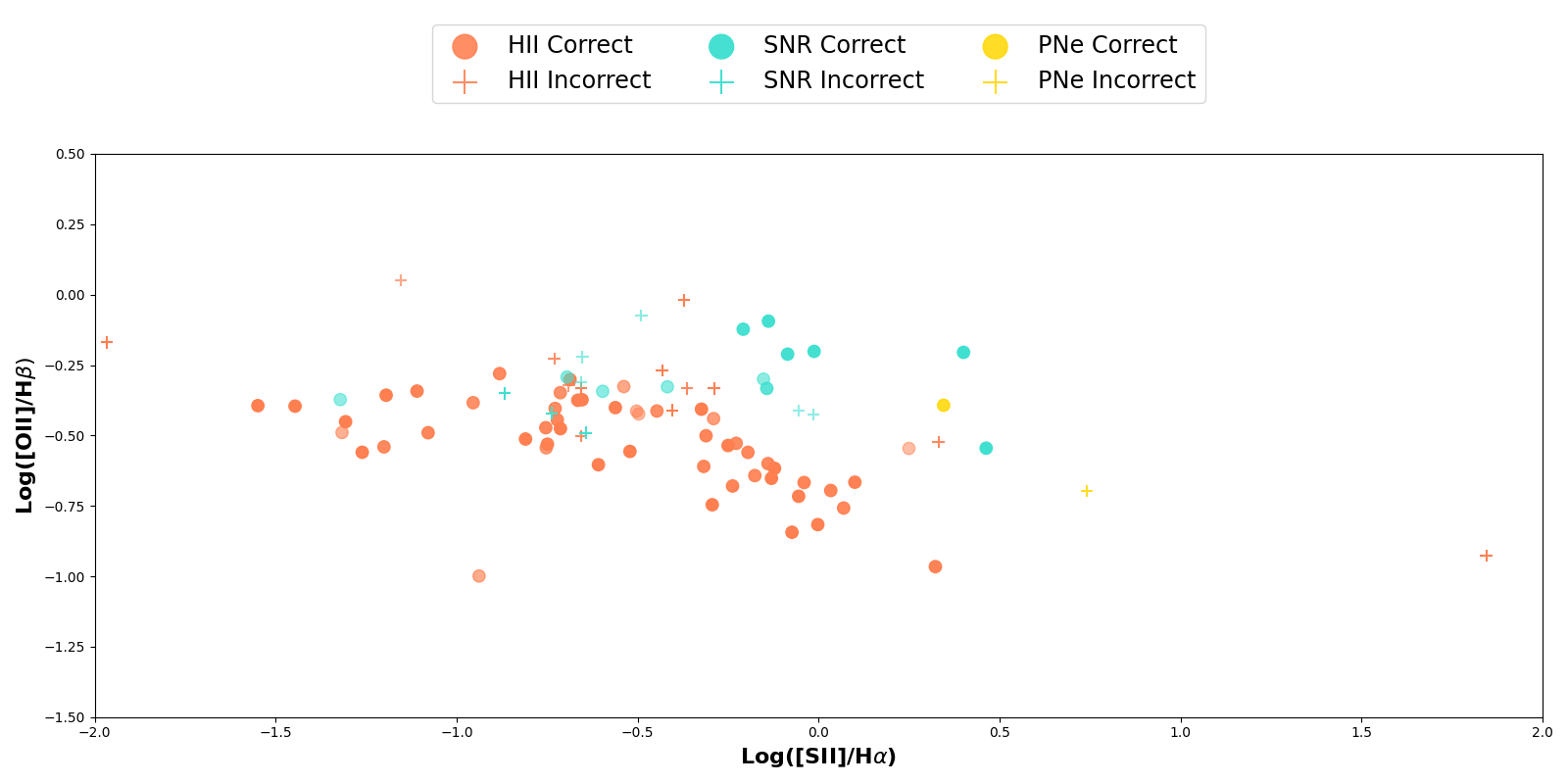}
    \caption{The line ratio diagram of log([{\sc O\,ii}]/H$\beta$) vs. log([{\sc S\,ii}]/H$\alpha$)  for the regions identified in a SITELLE field of M33.  The emission region type is designated by differing colors. Additionally, the correct and incorrect classification (True and False in the legend) are indicated by circles and crosses, respectively. Note these classifications were made with the network taking four input variables instead of three. The points' opacity corresponds to the probability of the classification. The lowest probability value is approximately 60\%.} 
    \label{fig:M33-field7-kk2}
\end{figure*}
\subsection{The Results}
\label{results}

In order to demonstrate the applicability of our two network frameworks described previously, we first apply the threshold on planetary nebulae luminosities derived above -- log($L_{H_\alpha}$) = 36.0 $\pm$ 0.1. Regions with an H$\alpha$ luminosity below this threshold were passed to the NN including planetary nebula as an output classification; regions with a higher value were passed the NN restricted to \hii{} regions and supernova remnants. There were 93 regions above this cut and 7 below.
After experimentation, we found that on real data it is best to use all four input parameters if we have three classification options and only three input parameters (i.e. we do not use {\sc O\,ii}]$\lambda3726,3729$/[{\sc O\,iii}]$\lambda5007$) if we have two classification options.


Figure \ref{fig:M33-field7-confusion} shows the confusion matrix after running the network developed in $\S$\ref{sec:meth} on the \textbf{100} zones in Field 7 of M33. As illustrated in the figure, the network excels at categorizing \hii{} regions -- as expected by the results reported in $\S$ \ref{sec:results-cm}. 
The confusion matrix reveals that one planetary nebula is correctly categorized while the other is categorized as a supernova remnant; the spectra for these PNe are shown in figure \ref{fig:PNe-plots}.
Furthermore, the network correctly classifies supernova remnants for approximately 65\% of the regions (i.e. 14 regions of 22 total regions). The remaining supernova remnants are incorrectly classified as \hii{} regions. 
Therefore, we explore in detail the reasoning behind the network's miscategorizations of the supernova remnants.

Figures \ref{fig:M33-field7-kk} and \ref{fig:M33-field7-al} show the placement of the M33 emission regions on standard BPT diagrams. From the figures, it is clear that the supernova remnants lie within the same regions as the \hii{} regions in these diagrams. More importantly, by comparing Figures \ref{fig:M33-field7-kk} and \ref{fig:M33-field7-al} with \ref{fig:NIIvsOIII} and \ref{fig:SIIvsOIII},  we note that the supernova remnants in M33 do not fall within the same regions of line-ratio space as the training set; this explains why the network does a poor job of accurately classifying these regions. Figures \ref{fig:M33-field7-kk2} shows similarly the placements of the regions identified using a network trained with the 4 input parameters over the [{\sc O\,ii}]$\lambda3726,3729$/H$\beta$ vs [{\sc S\,ii}]$\lambda6717$/H$\alpha$ diagram. We note that the \hii regions with a high [{\sc S\,ii}]$\lambda6717$/H$\alpha$  ratio are wrongfully identified as SNR, and the SNR with a high [{\sc O\,ii}]$\lambda3726,3729$/H$\beta$ ratio are also generally wrongfully identified as \hii regions.

Recent studies show that the parameter space indicated by the M33 supernovae may in fact be spanned by old supernova remnants (e.g. \citealt{moumen_3d_2019}). 
Figure \ref{fig:M33} shows the location of the supernova remnants in cyan, the planetary nebulae in green, and the \hii{} regions in yellow. As we can see, the regions identified as supernova remnants in the literature do not correspond with regions of strong emission; rather, they correspond with regions that would be considered as diffuse \hii{} emission which is how our network classifies them.
Therefore, we conclude that the misclassification of SNRs is either due to contamination from nearby \hii{} regions or DIGs (e.g., \citealt{cid_fernandes_detection_2021}; see appendices \ref{app:27} and \ref{app:miss} for an in-depth discussion), misclassification of these regions in the literature, or possibly an incomplete training set for supernova remnants (i.e. since our training set is a grid of shock models instead of supernova remnant models). We stress that the most likely culprit for the misclassification is the contamination from surrounding regions or incorrect classifications in the literature. However, a more comprehensive suite of SNR simulations under these conditions would be useful in improving the accuracy of the network on real data.
Additionally, comparing the true and false classifications in these plots reveals that the network is automatically learning similar diagnostics as the ones presented by \cite{kewley_optical_2001}.

As noted in $\S$\ref{sec:noise}, there exist gaps in the supernova remnant data owing to the metalicity grid employed when creating the dataset. Coincidentally, several misclassified supernova remnants lie in these gaps. Therefore, in order to ensure that these misclassifications are not due to the gaps in the training set, we interpolate the grid in metallicity in order to construct a new dataset that does not present gaps (see appendix \ref{app:miss-interp}). Using this complete dataset, we reconstruct our training, validation, and test sets. We then retrain our network and apply it to the M33 sample. The results indicate that while supernova remnant classification slightly increases, several \hii{} regions are now misclassified as supernova remnants. These results highlight the importance of the training set when applying machine learning algorithms to real data.

\section{Conclusions}\label{sec:conclusions}
In this work, we have demonstrated the feasibility of using a popular machine learning paradigm, artificial neural networks, to classify extragalactic emission regions into three categories: classic \hii{} regions, planetary nebulae, and supernova remnants. Synthetic line ratios for each emission type are generated using line amplitudes taken from 3MdB. We train, validate, and test our network on approximately 90,000 synthetic data containing three key line ratios: ([{\sc O\,iii}]$\lambda5007$/H$\beta$, [{\sc N\,ii}]$\lambda6583$/H$\alpha$, ([{\sc S\,ii}]$\lambda6717$+[{\sc S\,ii}]$\lambda6731$)/H$\alpha$. We also test the addition of a fourth line ratio: [{\sc O\,ii}]$\lambda3726$/H$\beta$.
We demonstrate the observational identification efficiency of this method by applying it to the Southwest field of M33; our results corroborate with existing literature with the exception of supernova remnants. We stress that these results are either due to an inconsistency between the training set and the real data, a common occurrence in supervised machine learning problems, incorrect identifications in the literature, or contamination from surrounding \hii{} or DIG. In order to resolve this inconsistency, a training set that explores the same region of parameter space for SNRs is required; we note this currently does not exist. 
Although the method was created with SITELLE in mind, the results can readily be ported to other instruments as long as the line fluxes are recovered. 

Additionally, this work highlights an important caveat that must be taken into account when applying machine learning algorithms to real astronomical data: without rigorous testing on real data, the results of the algorithm should be taken lightly. That is to say that it is inadvisable to use a machine learning algorithm that has not been verified  thoroughly on real data to a new data set. As demonstrated here, even though the network does an excellent job (95\%) classifying supernova remnants, it fails to do so correctly in real data due to the issues described above. 

Moreover, this work serves to expand the usage of machine learning in astronomy for classification purposes. Although we focused on three key categories of extragalactic emission line regions, this work can be extended to classify other objects in large catalogs of emission line regions (see figure \ref{fig:BPTDecision}). 

Another important conclusion of this work remains the inherent dependency of our method to the photoionization models' ability to reproduce properly the natural properties and physics of the PNe, SNR and \hii{} regions. The photoionization conditions, including the wide varieties of ionizing sources, the morphology and characteristics of their surrounding ISM, and the accurate modeling of the physical interactions between them, need to be fully understood while constraining their space of parameters covered in Nature. Surveys of ionized regions in all conditions and at a high spatial resolution that include a good characterization of ionizing sources can help to constrain the parameters while testing the photoionization models' performances.

Moreover, these large surveys, such as SIGNALS (\citealt{rousseau-nepton_signals_2019}) will enable us to amass a large quantity of well-resolved \hii{} and supernova remnants. Since we will have an adequate spectral resolution from these surveys, precise flux measurements can be made. Overall, these surveys will enable to the creation of a training set based on real observations instead of simulations. Although bias will always remain in any training set, whether with simulated or real data, this approach is expected to yield more accurate results. Furthermore, combining these large surveys with multiwavelength observations can ensure the correct classification of regions.

\section*{Acknowledgements}

The authors would like to thank the Canada-France-Hawaii Telescope (CFHT) which is operated by the National Research Council (NRC) of Canada, the Institut National des Sciences de l'Univers of the Centre National de la Recherche Scientifique (CNRS) of France, and the University of Hawaii. The observations at the CFHT were performed with care and respect from the summit of Maunakea which is a significant cultural and historic site.
C.L. R. acknowledges financial support from the physics department of the Universit\'e de Montr\'eal, the MITACS scholarship program, and the IVADO doctoral excellence scholarship.
J. H.-L. acknowledges support from NSERC via the Discovery grant program, as well as the Canada Research Chair program.
NVA acknowledges the support of the Royal Society and the Newton Fund via the award of a Royal Society--Newton Advanced Fellowship (grant NAF\textbackslash{}R1\textbackslash{}180403), and of Funda\c{c}\~ao de Amparo \`a Pesquisa e Inova\c{c}\~ao de Santa Catarina (FAPESC) and Conselho Nacional de Desenvolvimento Cient\'{i}fico e Tecnol\'{o}gico (CNPq).
CR is grateful to the Fonds de recherche du Québec - Nature et Technologies (FRQNT), for SIGNALS team financial support, and to the Natural Sciences and Engineering Research Council of Canada (NSERC)
K.G. is supported by the Australian Research Council through the Discovery Early Career Researcher Award (DECRA) Fellowship DE220100766 funded by the Australian Government. 
K.G. is supported by the Australian Research Council Centre of Excellence for All Sky Astrophysics in 3 Dimensions (ASTRO~3D), through project number CE170100013. 
The research of L. Chemin is funded by the Fondecyt Regular project 1210992 from Agencia Nacional de Investigacion y Desarrollo de Chile.

\section*{Data Availability}

 All data and methods used in this paper are available at 
 \href{https://github.com/sitelle-signals/Pamplemousse}{\faicon{github} sitelle-signals:Pamplemousse}.

\subsection*{Software}
python \citep{van_rossum_python_2009}, numpy \citep{van_der_walt_numpy_2011}, scipy \citep{virtanen_scipy_2020}, matplotlib \citep{hunter_matplotlib_2007}, pandas \citep{mckinney_data_2010}, seaborn \citep{michael_waskom_mwaskomseaborn_2017}, \citep{robitaille_astropy_2013}, tensorflow \citep{abadi_tensorflow_2015}, keras \citep{chollet_keras_2015}, \texttt{LUCI} \citep{rhea_luci_2021}.



\bibliographystyle{mnras}
\bibliography{BPT} 




\appendix

\section{Criterion for Filtering Data}
In this section we describe the filters applied to the 3MdB dataset for reproducibility's sake. The python application of these filters can be found in the notebook entitled \texttt{generating\_data.ipynb}.
To filter the \hii{} regions, we keep \texttt{lU\_mean} values equal to -2.5, -3.0, and -3.5. We only retain \texttt{fr==3.0}. We only keep \texttt{ab\_0} values between -5.4 and -3.0. 
The only additional filtering we apply on the PNe dataset is \texttt{com6==1}. For the shock models, we only take \texttt{emis\_VI.model\_type='shock'} and \texttt{abundances.name='Allen2008\_Solar'}.

\section{Diagnostic Plots of Test Set}
In this section we present the diagnostic plots of the test set data (as shown in figures \ref{fig:NIIvsOIII} and \ref{fig:SIIvsOIII}) broken into each type.

\begin{figure*}
    \centering
        \begin{subfigure}{0.33\textwidth}
            \includegraphics[width=\textwidth]{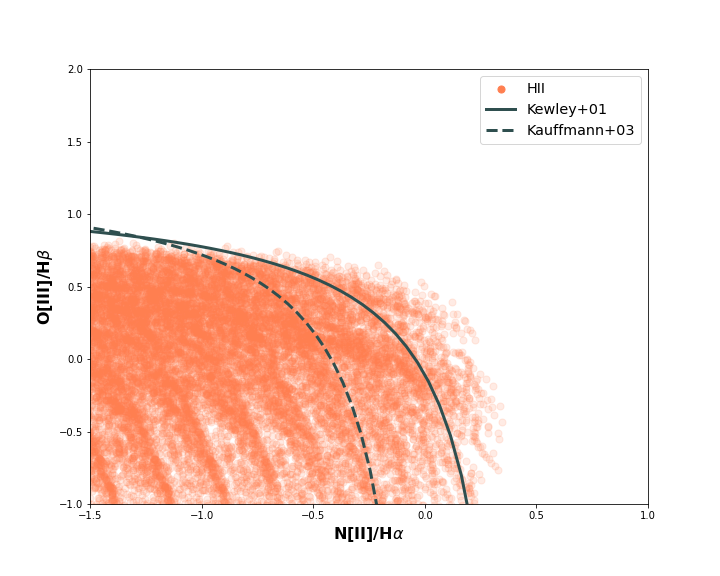}
            \caption{\hii{} Regions}
        \end{subfigure}
        \hfill
        \begin{subfigure}{0.33\textwidth}
            \includegraphics[width=\textwidth]{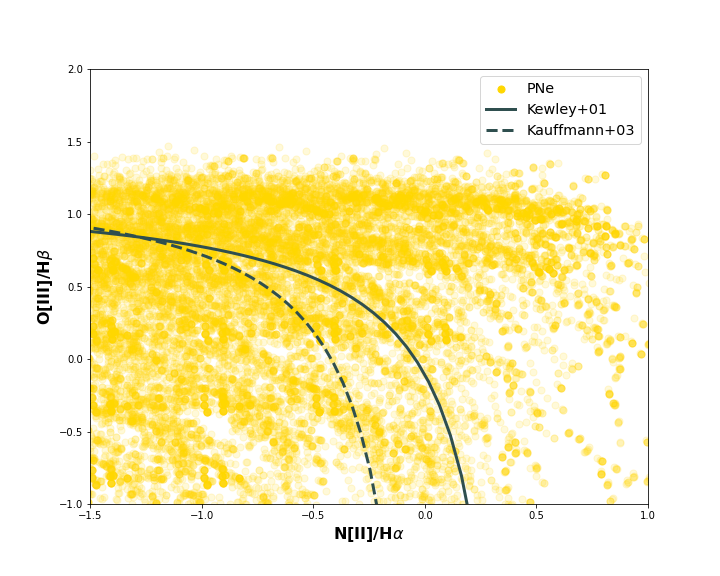}
            \caption{Planetary Nebulae}
        \end{subfigure}
        \hfill
        \begin{subfigure}{0.33\textwidth}            
            \includegraphics[width=\textwidth]{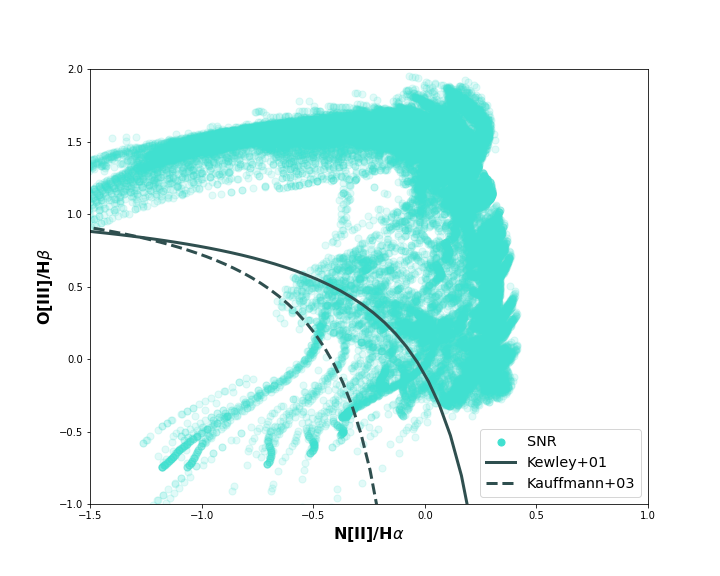}
            \caption{Supernova Remnants}
        \end{subfigure}
    \caption{\nii{}/H$\alpha$ vs. \oiii{}/H$\beta$ Diagnostic plot broken down by emission region type.}
    \label{fig:BPT1}
\end{figure*}

\begin{figure*}
    \begin{subfigure}{0.33\textwidth}
    \includegraphics[width=\textwidth]{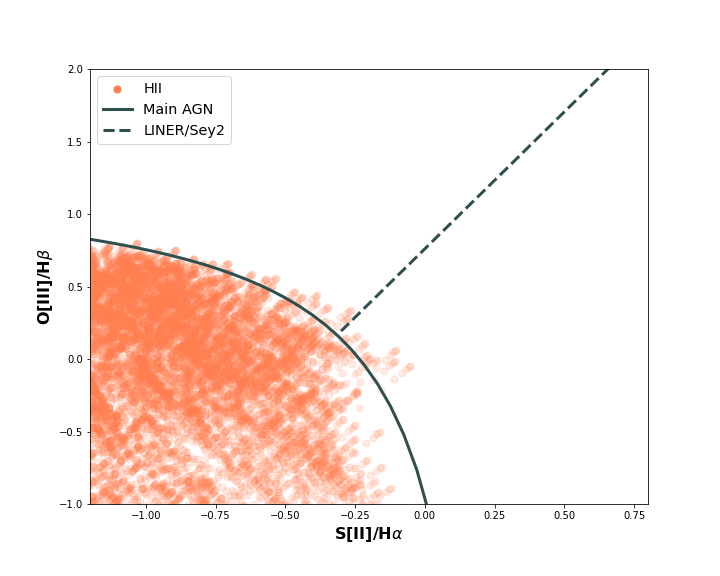}
    \caption{\hii{} Regions}
    \end{subfigure}
    \hfill
    \begin{subfigure}{0.33\textwidth}
    \includegraphics[width=\textwidth]{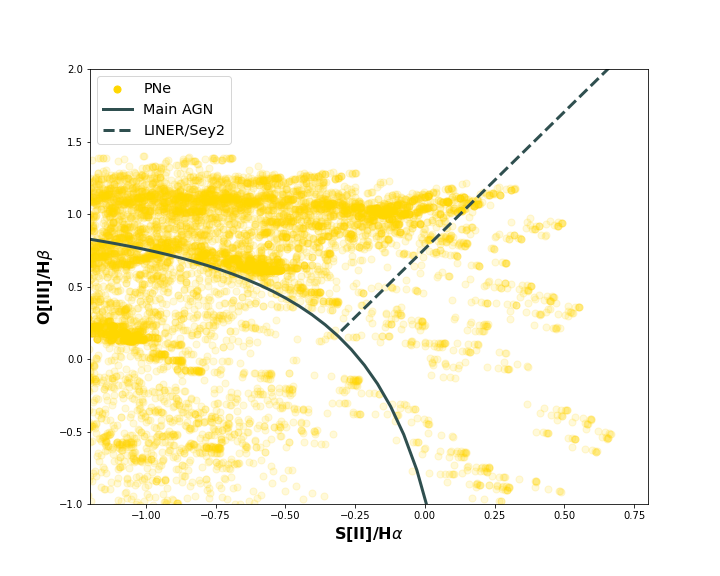}
    \caption{Planetary Nebulae}
    \end{subfigure}
    \hfill
    \begin{subfigure}{0.33\textwidth}
    \includegraphics[width=\textwidth]{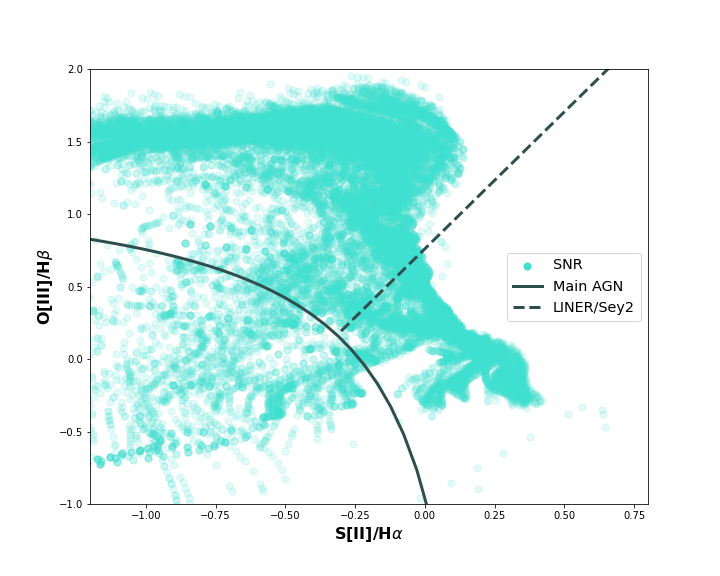}
    \caption{Supernova Remnants}
    \end{subfigure}
    \caption{\sii{}/H$\alpha$ vs. \oiii{}/H$\beta$ Diagnostic plot broken down by emission region type.}
    \label{fig:BPT2}
\end{figure*}

\section{Fit plots}
\begin{figure*}
    \begin{subfigure}{0.33\textwidth}
            \includegraphics[width=\textwidth]{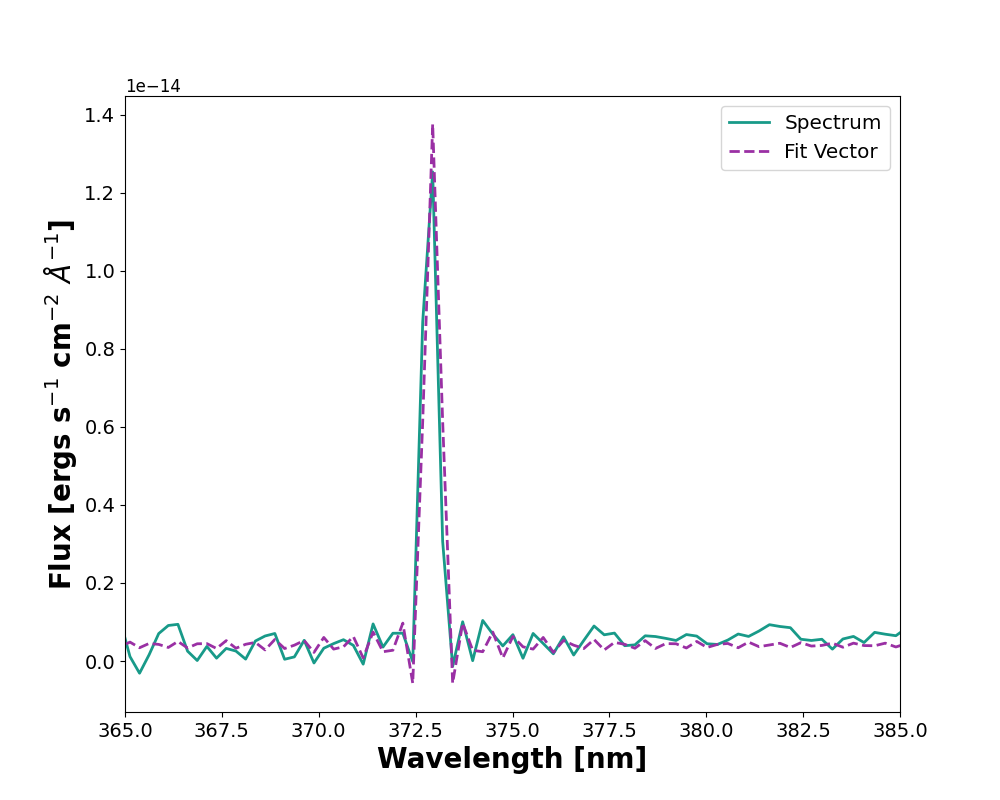}
            \caption{SN1 spectrum and fitted vector for correctly categorized \hii{} region}
            \hfill
            \end{subfigure}
            \begin{subfigure}{0.33\textwidth}
            \includegraphics[width=\textwidth]{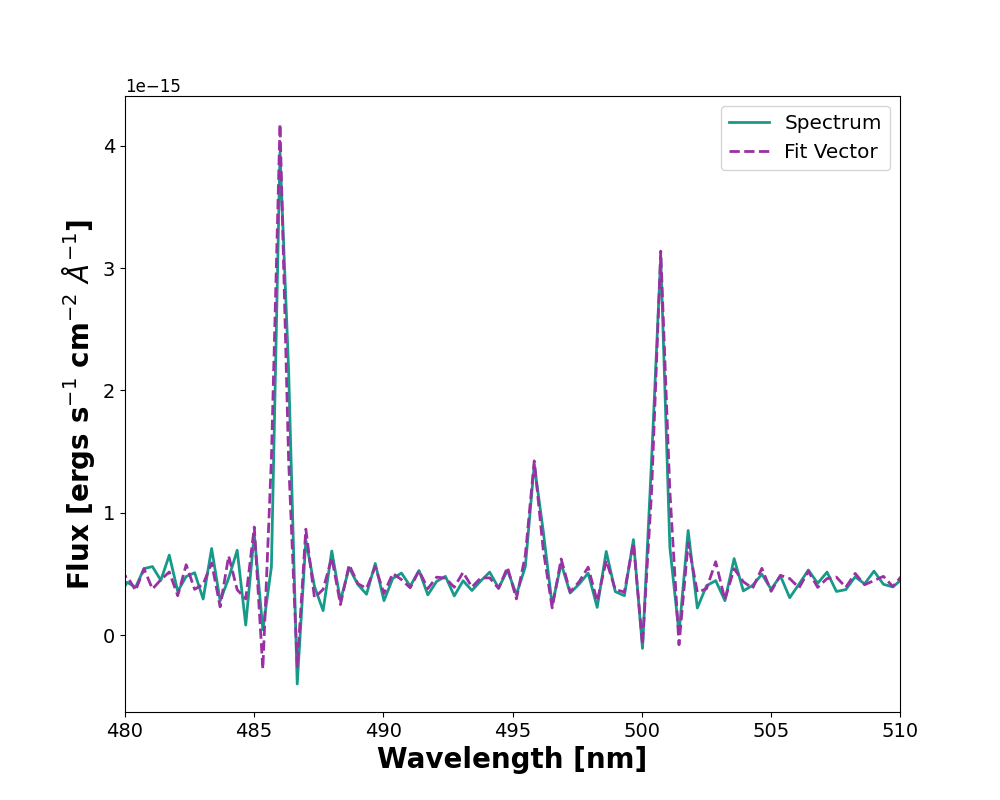}
            \caption{SN2 spectrum and fitted vector for correctly categorized \hii{} region}
            \hfill
            \end{subfigure}
            \begin{subfigure}{0.33\textwidth}
            \includegraphics[width=\textwidth]{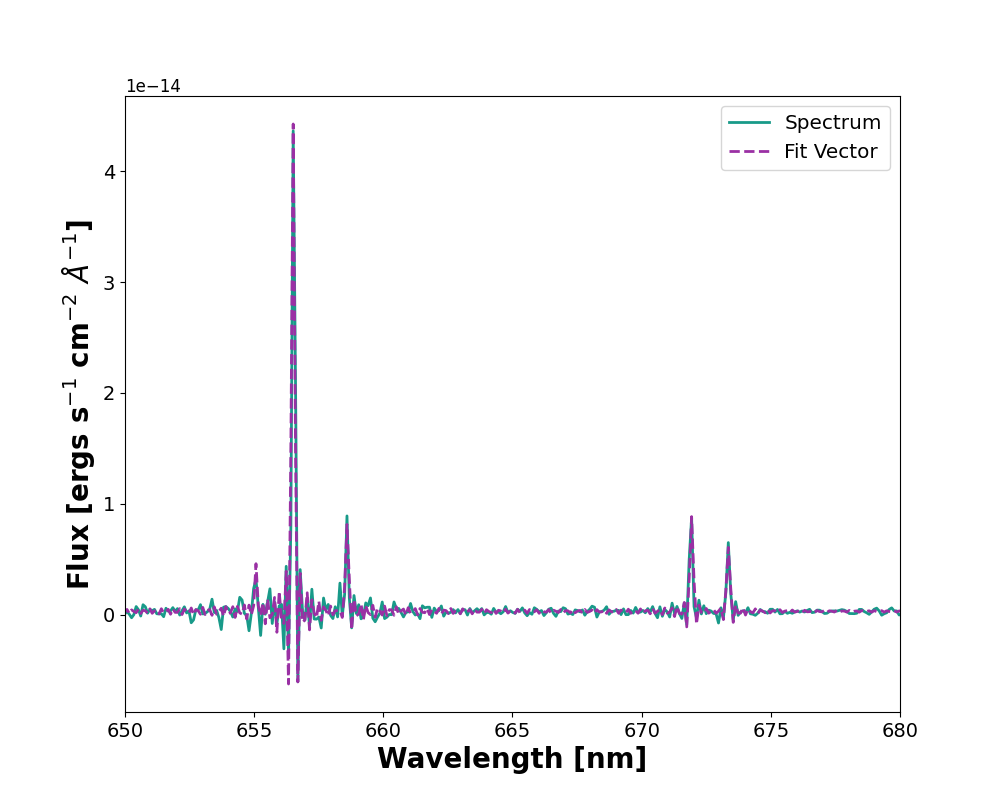}
            \caption{SN3 spectrum and fitted vector for correctly categorized \hii{} region}
            \end{subfigure}

            \begin{subfigure}{0.33\textwidth}
            \includegraphics[width=\textwidth]{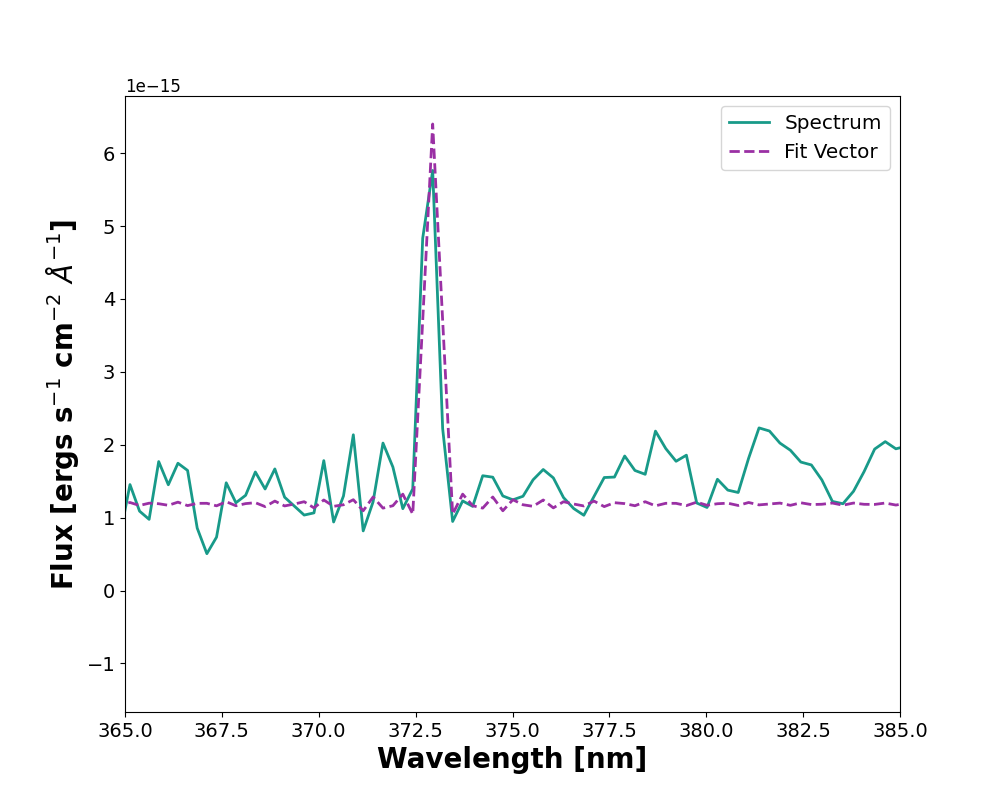}
            \caption{SN1 spectrum and fitted vector for incorrectly categorized \hii{} region}
             \end{subfigure}
            \begin{subfigure}{0.33\textwidth}
            \includegraphics[width=\textwidth]{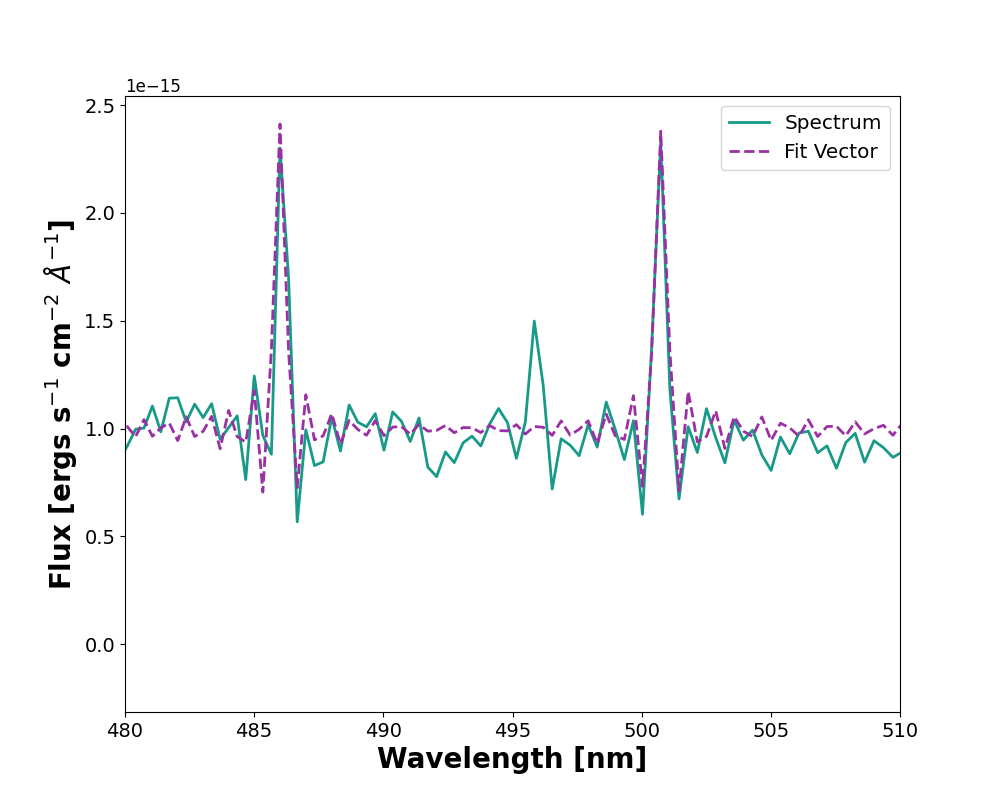}
            \caption{SN2 spectrum and fitted vector for incorrectly categorized \hii{} region}
             \end{subfigure}
            \begin{subfigure}{0.33\textwidth}
            \includegraphics[width=\textwidth]{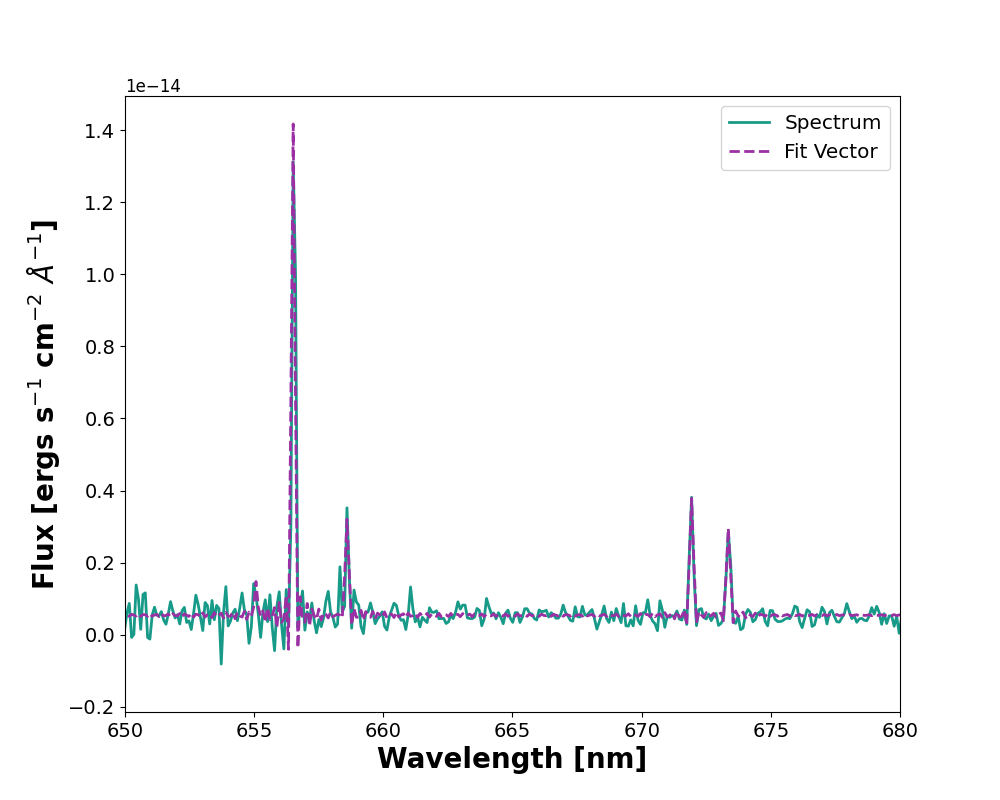}
            \caption{SN3 spectrum and fitted vector for incorrectly categorized \hii{} region}
             \end{subfigure}
    \caption{SN1, SN2, \& SN3 of correctly categorized \hii{} regions}
    \label{fig:hii-plots}
\end{figure*}

\begin{figure*}
    \begin{subfigure}{0.33\textwidth}
    \includegraphics[width=\textwidth]{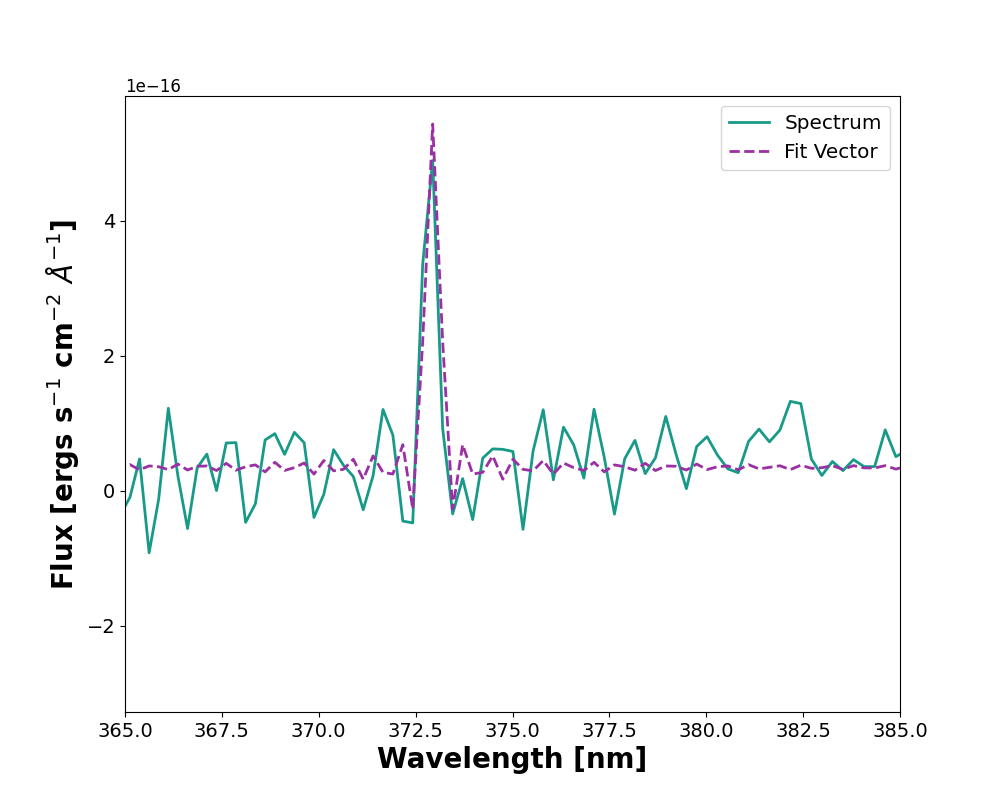}
     \caption{SN1 spectrum and fitted vector for correctly categorized PNe region}
     \hfill
     \end{subfigure}
     \begin{subfigure}{0.33\textwidth}
    \includegraphics[width=\textwidth]{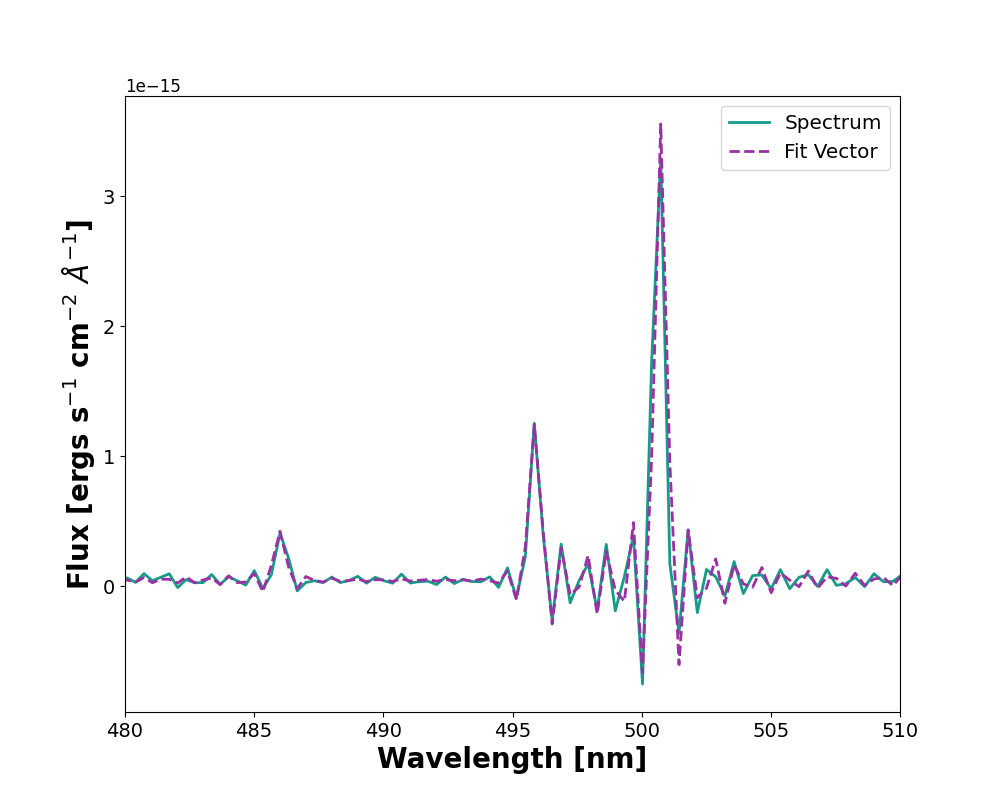}
     \caption{SN2 spectrum and fitted vector for correctly categorized PNe region}
     \hfill
     \end{subfigure}
     \begin{subfigure}{0.33\textwidth}
    \includegraphics[width=\textwidth]{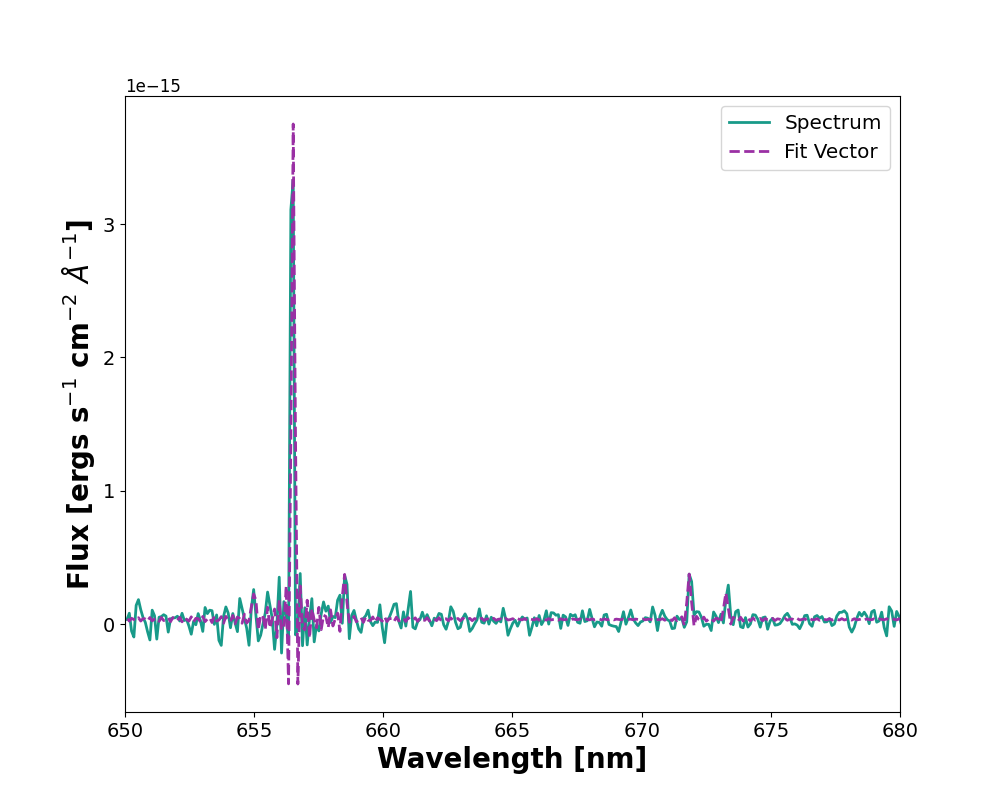}
     \caption{SN3 spectrum and fitted vector for correctly categorized PNe region}
     \hfill
     \end{subfigure}
     \begin{subfigure}{0.33\textwidth}
    \includegraphics[width=\textwidth]{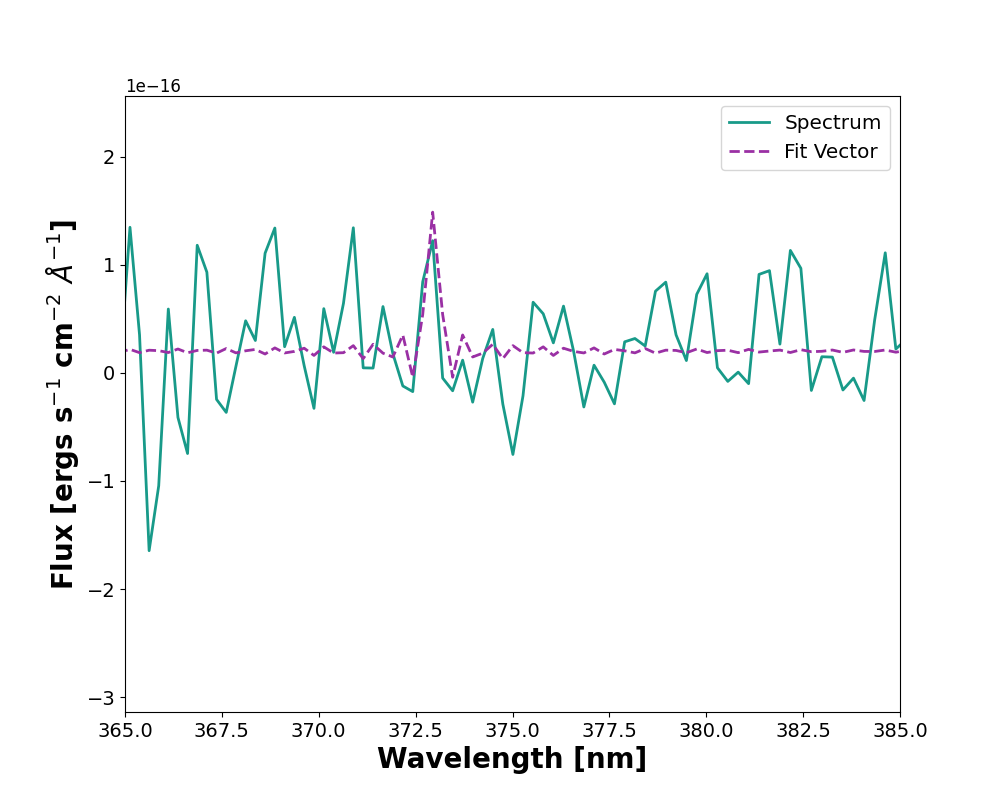}
    \caption{SN1 spectrum and fitted vector for incorrectly categorized PNe region}
     \hfill
     \end{subfigure}
     \begin{subfigure}{0.33\textwidth}
    \includegraphics[width=\textwidth]{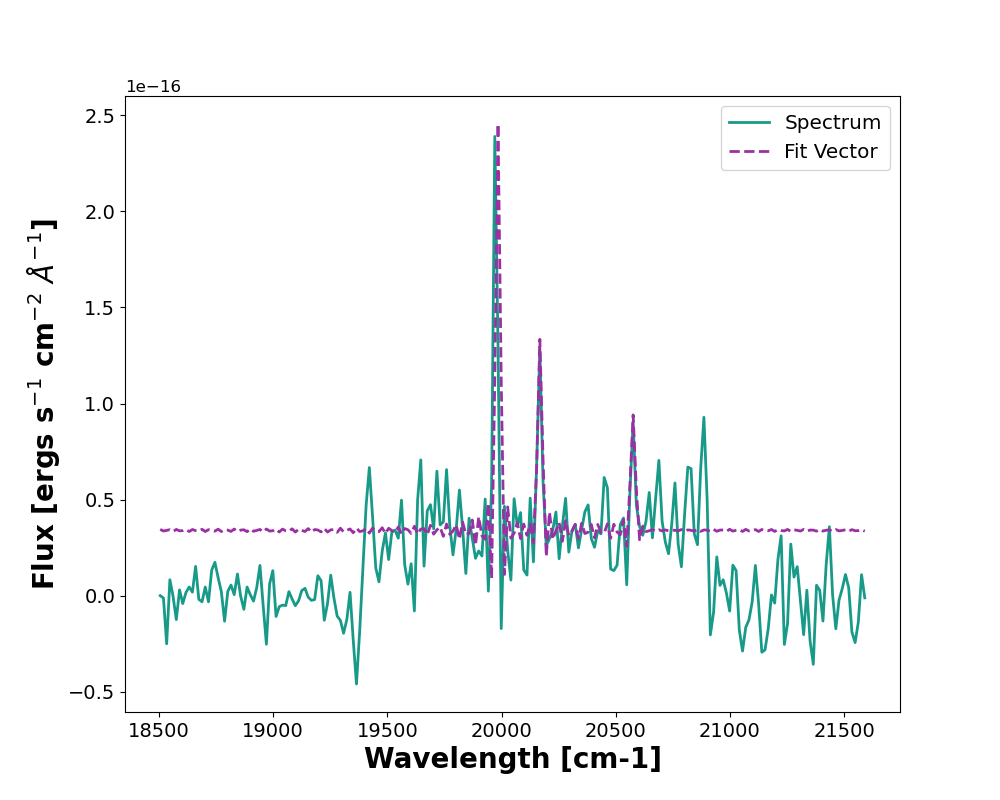}
    \caption{SN2 spectrum and fitted vector for incorrectly categorized PNe region}
     \hfill
     \end{subfigure}
     \begin{subfigure}{0.33\textwidth}
    \includegraphics[width=\textwidth]{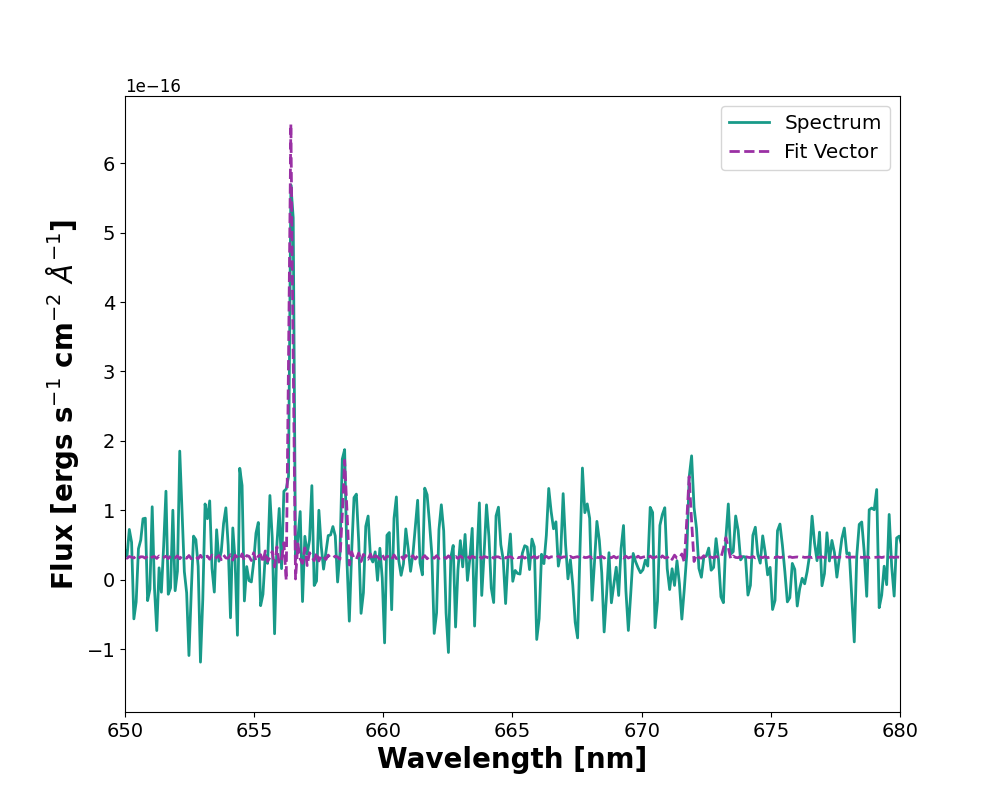}
    \caption{SN3 spectrum and fitted vector for incorrectly categorized PNe region}
     \hfill
     \end{subfigure}
            
    \caption{SN1, SN2, \& SN3 of the correctly and incorrectly categorized PNe regions. The region incorrectly classified was classified as a supernova remnant.
    }
    \label{fig:PNe-plots}
\end{figure*}

\begin{figure*}
    \begin{subfigure}{0.33\textwidth}
    \includegraphics[width=\textwidth]{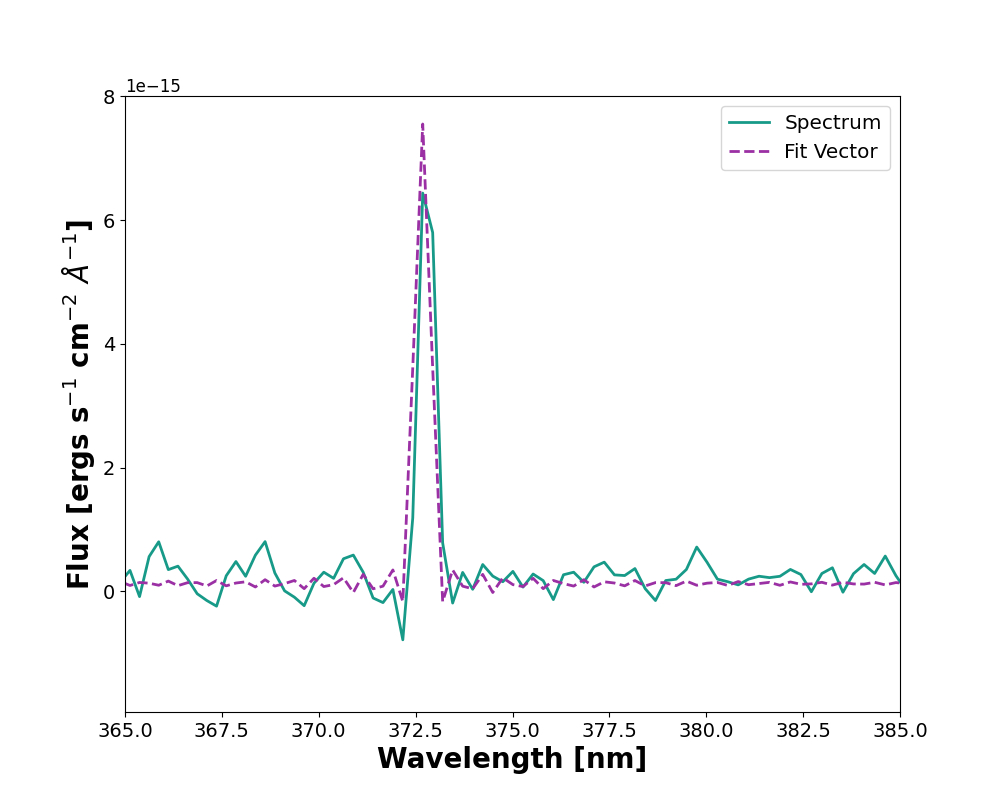}
    \caption{SN1 spectrum and fitted vector for correctly categorized SNR region}
    \hfill
    \end{subfigure}
    \begin{subfigure}{0.33\textwidth}
    \includegraphics[width=\textwidth]{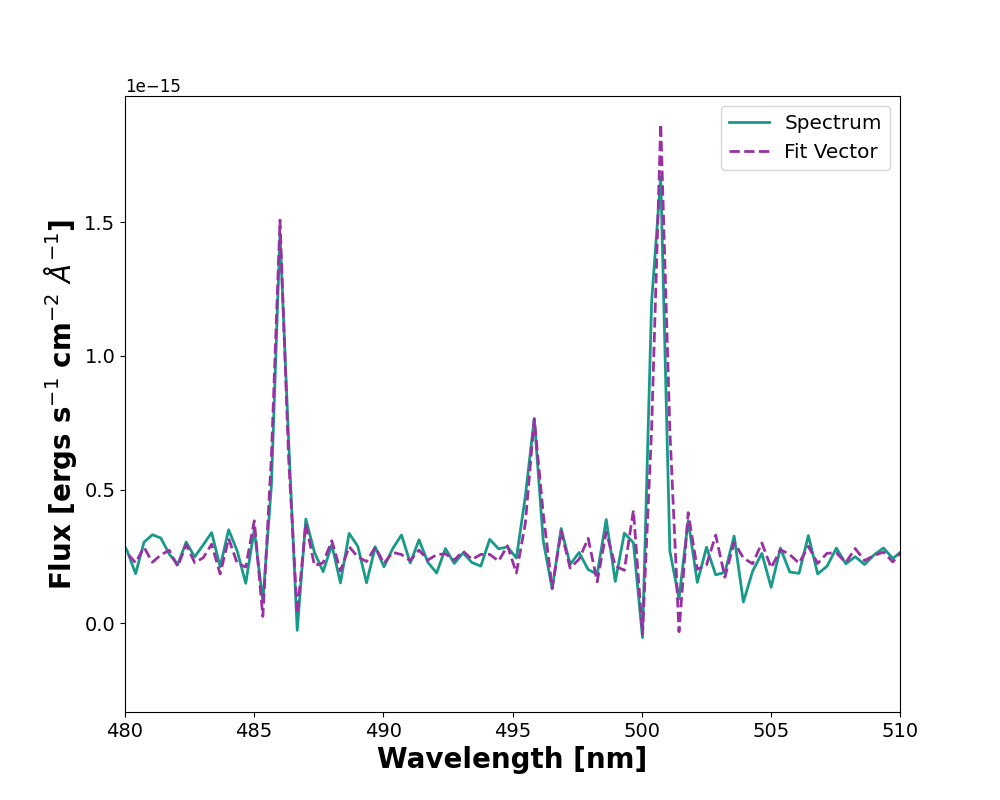}
    \caption{SN2 spectrum and fitted vector for correctly categorized SNR region}
    \hfill
    \end{subfigure}
    \begin{subfigure}{0.33\textwidth}
    \includegraphics[width=\textwidth]{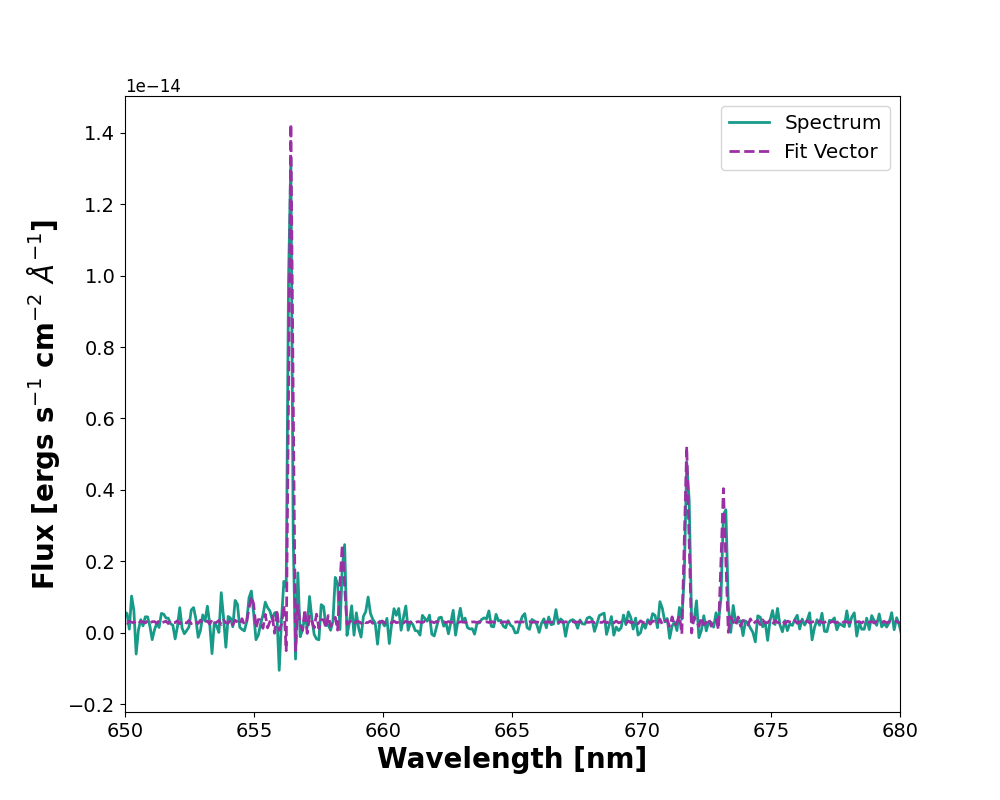}
    \caption{SN3 spectrum and fitted vector for correctly categorized SNR region}
    \hfill
    \end{subfigure}
    \begin{subfigure}{0.33\textwidth}
    \includegraphics[width=\textwidth]{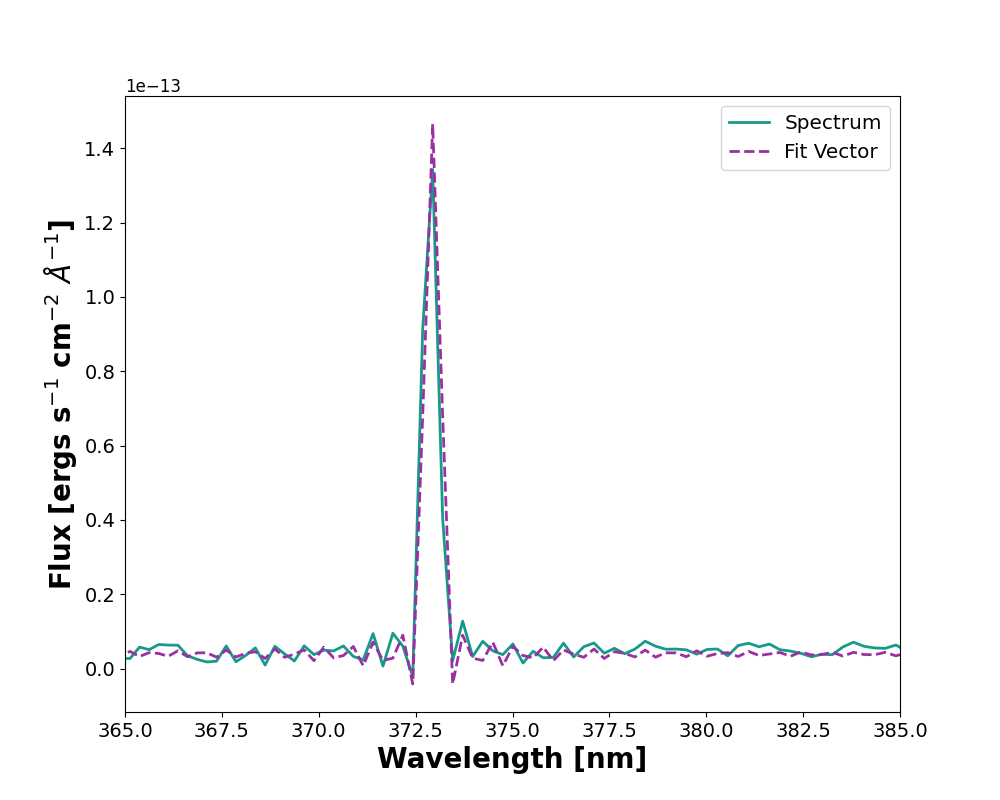}
    \caption{SN1 spectrum and fitted vector for incorrectly categorized SNR region}
    \hfill
    \end{subfigure}
    \begin{subfigure}{0.33\textwidth}
    \includegraphics[width=\textwidth]{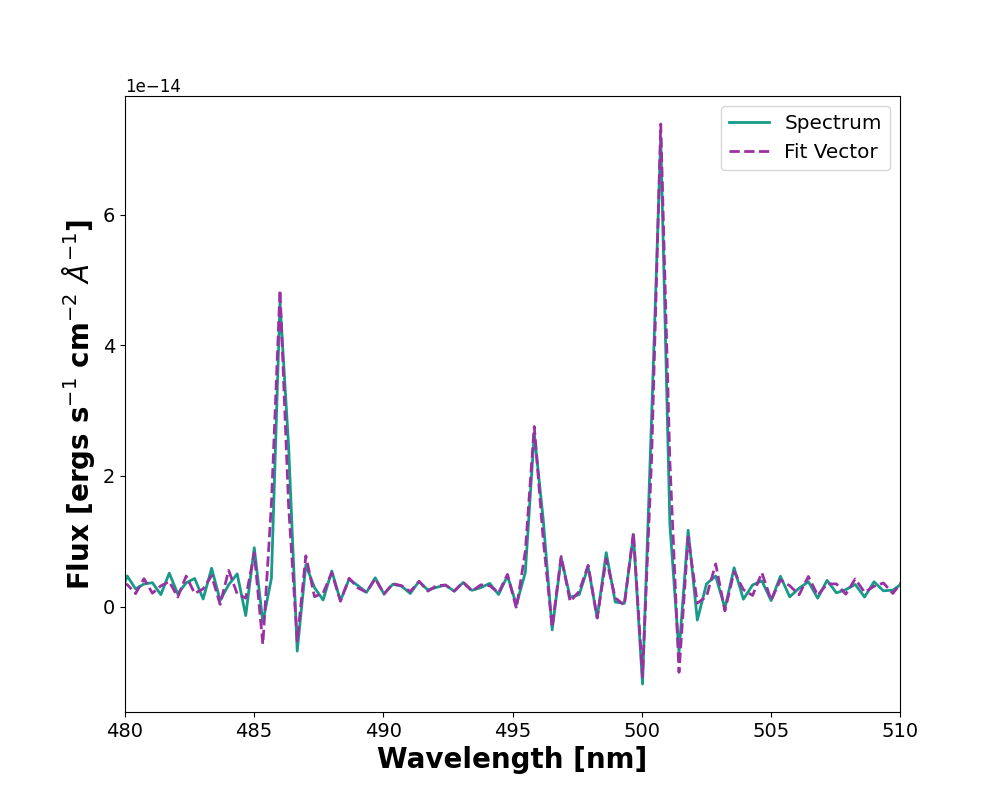}
    \caption{SN2 spectrum and fitted vector for incorrectly categorized SNR region}
    \hfill
    \end{subfigure}
    \begin{subfigure}{0.33\textwidth}
    \includegraphics[width=\textwidth]{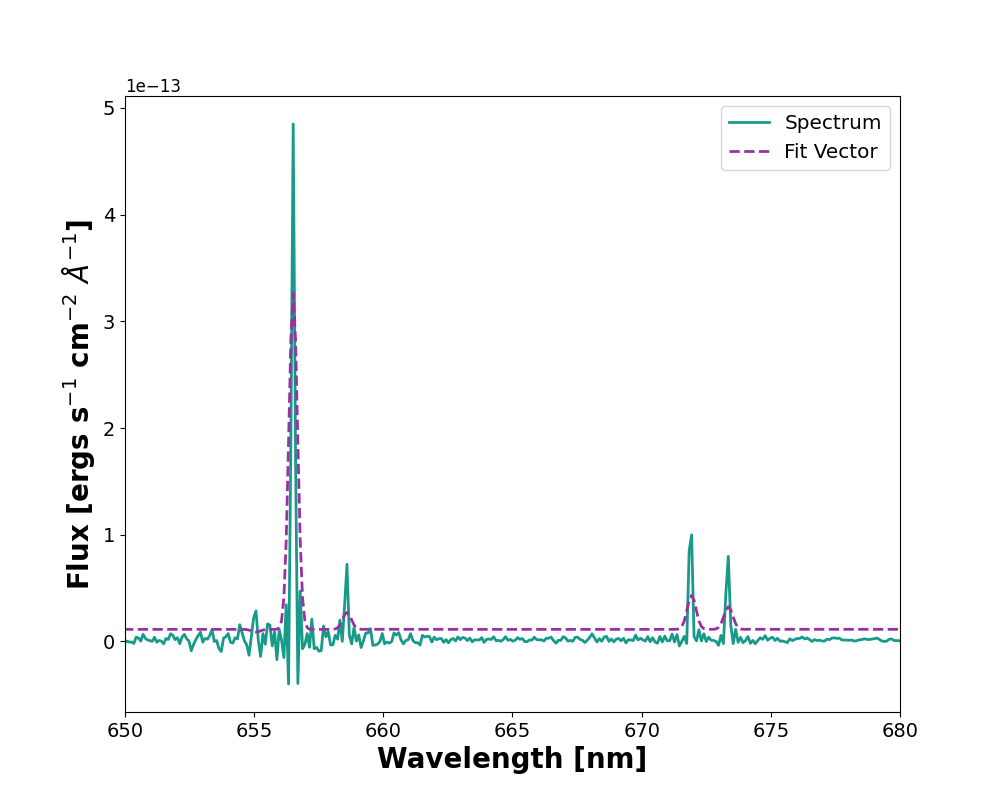}
    \caption{SN3 spectrum and fitted vector for incorrectly categorized SNR region}
    \hfill
     \end{subfigure}
    \caption{SN1, SN2, \& SN3 of correctly and incorrectly categorized SNR regions. The region incorrectly classified was classified as an \hii{} region.}
    \label{fig:SNR-plots}
\end{figure*}

This section contains three plots families (figure \ref{fig:hii-plots}, figure \ref{fig:PNe-plots}, and figure \ref{fig:SNR-plots}). Each family consists of the SN1, SN2, and SN3 fits of a emission region type (i.e. \hii{}, PNe, or SNR). Additionally, each family is further broken into the spectra which we correctly categorized and those incorrectly categorized.


\newpage
\section{List of Regions}
This section contains a table (Table \ref{tab:location}) documenting the regions of M33 used in this study. The table includes the region type, the RA, the DEC, and radius. The table only reports regions that were not rejected on the basis of insufficient signal-to-noise of their spectra.
\begin{table}
	\centering
	\caption{Table of regions, their RA, DEC, and radius in arcseconds. The regions are shown in figure \ref{fig:M33}.}
	\label{tab:location}
	\begin{tabular}{cccc} 
		\hline
		Region ID & RA & DEC & Radius \\
		\hline
        0   & 1:33:14.9493 & +30:32:29.863 & 10.440" \\ \hline
        2   & 1:33:12.2492 & +30:30:24.375 & 5.606"  \\ \hline
        3   & 1:33:11.9795 & +30:30:11.208 & 8.822"  \\ \hline
        4   & 1:33:13.8888 & +30:29:45.757 & 5.438"  \\ \hline
        5   & 1:33:11.2340 & +30:29:52.776 & 3.603"  \\ \hline
        6   & 1:33:09.6747 & +30:29:45.723 & 8.170"  \\ \hline
        7   & 1:33:10.0807 & +30:29:54.507 & 4.545"  \\ \hline
        8   & 1:33:12.4954 & +30:34:08.952 & 10.957" \\ \hline
        9   & 1:33:11.7219 & +30:38:56.571 & 34.026" \\ \hline
        10  & 1:33:13.5530 & +30:39:32.587 & 2.502"  \\ \hline
        11  & 1:33:13.6216 & +30:39:28.201 & 2.947"  \\ \hline
        15  & 1:33:06.1099 & +30:31:02.774 & 2.000"  \\ \hline
        16  & 1:33:08.5500 & +30:34:38.500 & 2.000"  \\ \hline
        22  & 1:33:11.0818 & +30:34:22.629 & 16.226" \\ \hline
        24  & 1:33:12.4386 & +30:38:43.268 & 10.000" \\ \hline
        25  & 1:33:14.1437 & +30:39:48.311 & 12.684" \\ \hline
        26  & 1:32:31.4769 & +30:35:32.901 & 13.528" \\ \hline
        27  & 1:32:46.7300 & +30:34:37.800 & 7.099"  \\ \hline
        29  & 1:32:53.3331 & +30:37:56.037 & 7.449"  \\ \hline
        30  & 1:32:57.6721 & +30:39:27.962 & 6.705"  \\ \hline
        31  & 1:33:00.1500 & +30:30:46.200 & 11.272" \\ \hline
        32  & 1:33:00.6700 & +30:30:59.300 & 6.911"  \\ \hline
        33  & 1:33:01.5100 & +30:30:47.874 & 3.977"  \\ \hline
        34  & 1:33:02.9300 & +30:32:28.737 & 8.102"  \\ \hline
        35  & 1:33:03.5700 & +30:31:20.037 & 5.971"  \\ \hline
        36  & 1:33:04.0971 & +30:39:58.016 & 8.697"  \\ \hline
        37  & 1:33:09.9369 & +30:39:34.899 & 12.389" \\ \hline
        38  & 1:33:11.1669 & +30:39:43.699 & 12.952" \\ \hline
        39  & 1:32:27.6494 & +30:35:44.598 & 13.786" \\ \hline
        40  & 1:32:35.4269 & +30:35:19.800 & 15.386" \\ \hline
        41  & 1:32:40.8064 & +30:31:51.099 & 16.877" \\ \hline
        42  & 1:32:42.7100 & +30:36:19.237 & 9.503"  \\ \hline
        43  & 1:32:52.9336 & +30:31:32.474 & 12.540" \\ \hline
        44  & 1:32:56.0532 & +30:33:30.400 & 12.923" \\ \hline
        45  & 1:32:57.2469 & +30:39:14.700 & 7.378"  \\ \hline
        54  & 1:32:45.6879 & +30:38:55.637 & 22.822" \\ \hline
        55  & 1:32:43.4136 & +30:38:53.039 & 8.238"  \\ \hline
        56  & 1:32:56.9272 & +30:40:18.512 & 8.238"  \\ \hline
        57  & 1:33:02.3461 & +30:39:53.458 & 4.602"  \\ \hline
        58  & 1:33:01.4757 & +30:39:29.292 & 4.602"  \\ \hline
        59  & 1:33:00.2709 & +30:39:02.536 & 5.676"  \\ \hline
        \end{tabular}
\end{table}
\begin{table}
	\centering
	\contcaption{table continued from the previous one.}
	\label{tab:example_table}
	\begin{tabular}{lccr} 
		\hline
        60  & 1:33:07.2179 & +30:35:12.876 & 5.583"  \\ \hline
        61  & 1:33:07.8867 & +30:35:18.914 & 4.481"  \\ \hline
        62  & 1:32:58.6626 & +30:36:37.518 & 5.583"  \\ \hline
        63  & 1:32:52.3103 & +30:37:14.648 & 4.428"  \\ \hline
        64  & 1:32:54.1829 & +30:37:28.459 & 2.852"  \\ \hline
        65  & 1:32:57.9253 & +30:34:50.479 & 5.583"  \\ \hline
        66  & 1:32:57.7246 & +30:34:41.847 & 5.583"  \\ \hline
        67  & 1:32:58.1255 & +30:34:34.077 & 3.210"  \\ \hline
        68  & 1:32:52.7781 & +30:34:55.667 & 11.849" \\ \hline
        69  & 1:32:56.5215 & +30:34:51.346 & 5.276"  \\ \hline
        70  & 1:32:52.9121 & +30:36:36.665 & 31.885" \\ \hline
        71  & 1:32:59.5295 & +30:34:43.568 & 8.761"  \\ \hline
        72  & 1:33:00.8657 & +30:34:18.529 & 11.849" \\ \hline
        73  & 1:33:00.9989 & +30:34:00.400 & 5.126"  \\ \hline
        74  & 1:33:00.5310 & +30:33:55.223 & 1.825"  \\ \hline
        75  & 1:33:04.4752 & +30:34:21.100 & 3.567"  \\ \hline
        76  & 1:33:03.8734 & +30:34:15.061 & 3.567"  \\ \hline
        77  & 1:33:04.4747 & +30:34:07.288 & 6.463"  \\ \hline
        78  & 1:33:02.6696 & +30:33:50.897 & 3.567"  \\ \hline
        79  & 1:33:03.4707 & +30:33:22.406 & 6.862"  \\ \hline
        80  & 1:33:05.0067 & +30:32:54.773 & 10.142" \\ \hline
        81  & 1:33:02.8685 & +30:32:58.238 & 6.560"  \\ \hline
        82  & 1:32:56.9878 & +30:32:54.808 & 6.862"  \\ \hline
        83  & 1:32:51.7755 & +30:33:04.309 & 6.553"  \\ \hline
        84  & 1:32:55.7178 & +30:32:26.324 & 8.094"  \\ \hline
        86  & 1:33:04.7373 & +30:31:55.212 & 5.268"  \\ \hline
        87  & 1:33:06.2730 & +30:31:31.894 & 3.843"  \\ \hline
        88  & 1:33:06.6737 & +30:31:27.575 & 2.922"  \\ \hline
        90  & 1:33:07.6752 & +30:31:14.620 & 2.922"  \\ \hline
        91  & 1:33:07.6076 & +30:30:57.355 & 4.015"  \\ \hline
        92  & 1:33:06.8067 & +30:31:14.626 & 4.015"  \\ \hline
        93  & 1:33:04.8019 & +30:30:56.511 & 3.832"  \\ \hline
        94  & 1:33:04.5342 & +30:30:42.701 & 2.770"  \\ \hline
        95  & 1:33:08.2056 & +30:29:50.051 & 14.323" \\ \hline
        96  & 1:32:52.2431 & +30:29:37.996 & 37.088" \\ \hline
        97  & 1:32:45.3615 & +30:31:56.103 & 4.646"  \\ \hline
        98  & 1:32:38.6850 & +30:29:47.449 & 4.646"  \\ \hline
        99  & 1:32:34.6761 & +30:30:21.086 & 11.742" \\ \hline
        \end{tabular}
\end{table}
        \begin{table}
	\centering
	\contcaption{table continued from the previous one.}
	\label{tab:example_table}
	\begin{tabular}{lccr} 
        101 & 1:32:29.9250 & +30:32:36.569 & 11.742" \\ \hline
        102 & 1:32:31.5290 & +30:32:32.269 & 5.541"  \\ \hline
        103 & 1:32:46.6289 & +30:34:07.319 & 10.078" \\ \hline
        104 & 1:32:44.5559 & +30:34:53.927 & 7.709"  \\ \hline
        105 & 1:32:44.4886 & +30:35:16.371 & 3.318"  \\ \hline
        106 & 1:32:43.5526 & +30:35:16.367 & 3.341"  \\ \hline
        107 & 1:32:44.8228 & +30:35:17.235 & 2.441"  \\ \hline
        109 & 1:32:44.2201 & +30:36:02.985 & 7.709"  \\ \hline
        110 & 1:32:41.6795 & +30:36:01.247 & 4.735"  \\ \hline
        112 & 1:32:29.5105 & +30:36:08.058 & 18.767" \\ \hline
        113 & 1:32:31.1856 & +30:35:07.648 & 31.644" \\ \hline
        115 & 1:32:42.5477 & +30:36:34.918 & 3.341"  \\ \hline
        116 & 1:32:42.0797 & +30:36:34.916 & 3.341"  \\ \hline
        117 & 1:32:55.7223 & +30:39:30.173 & 3.590"  \\ \hline
        118 & 1:32:55.2541 & +30:39:34.490 & 4.428"  \\ \hline
        119 & 1:32:47.0250 & +30:39:47.435 & 2.393"  \\ \hline
        120 & 1:32:44.5502 & +30:39:20.667 & 2.166"  \\ \hline
        123 & 1:32:53.1800 & +30:38:55.646 & 2.517"  \\ \hline
        124 & 1:32:33.2680 & +30:32:01.209 & 3.485"  \\ \hline
        125 & 1:32:28.4494 & +30:33:59.424 & 2.922"  \\ \hline
        126 & 1:32:37.7140 & +30:39:54.586 & 29.031" \\ \hline
        128 & 1:33:00.9338 & +30:35:05.144 & 17.306" \\ \hline
		\hline
  
	\end{tabular}
\end{table}

\section{The strange case of ID 27}\label{app:27}
Region 27, the supernova remnant, was misclassified despite the fact that the combined SITELLE deep image clearly shows a SNR. While the other misclassifications of supernova remnants are attributable to other causes (see discussion in $\S$ \ref{sec:M33}),  the reason for this incorrect categorization is not clear without examining the spectrum in SN3; the \sii{} doublet and H$\alpha$ emission clearly demonstrate the presence of multiple components. Thus, the fluxes for these lines are incorrectly estimated which leads the network to classify it as an \hii{} region. The presence of multiple lines is likely due to contamination from the surrounding DIG or \hii{} region. We do not see these multiple components in the other filters since their spectral resolution is insufficient to resolve them.
Although an in depth examination of this region is beyond the scope of this paper, we have attempted to quantify this effect in the following manner: 1) we fit two H$\alpha$ and \sii{} lines and 2) we artificially reduce the spectral resolution to 3000. Although we do not repeat this process for the 7 incorrectly categorized SNRs that exhibit multiple components, region 27 serves as an example.

\begin{figure}
    \centering
    \includegraphics[width=0.45\textwidth]{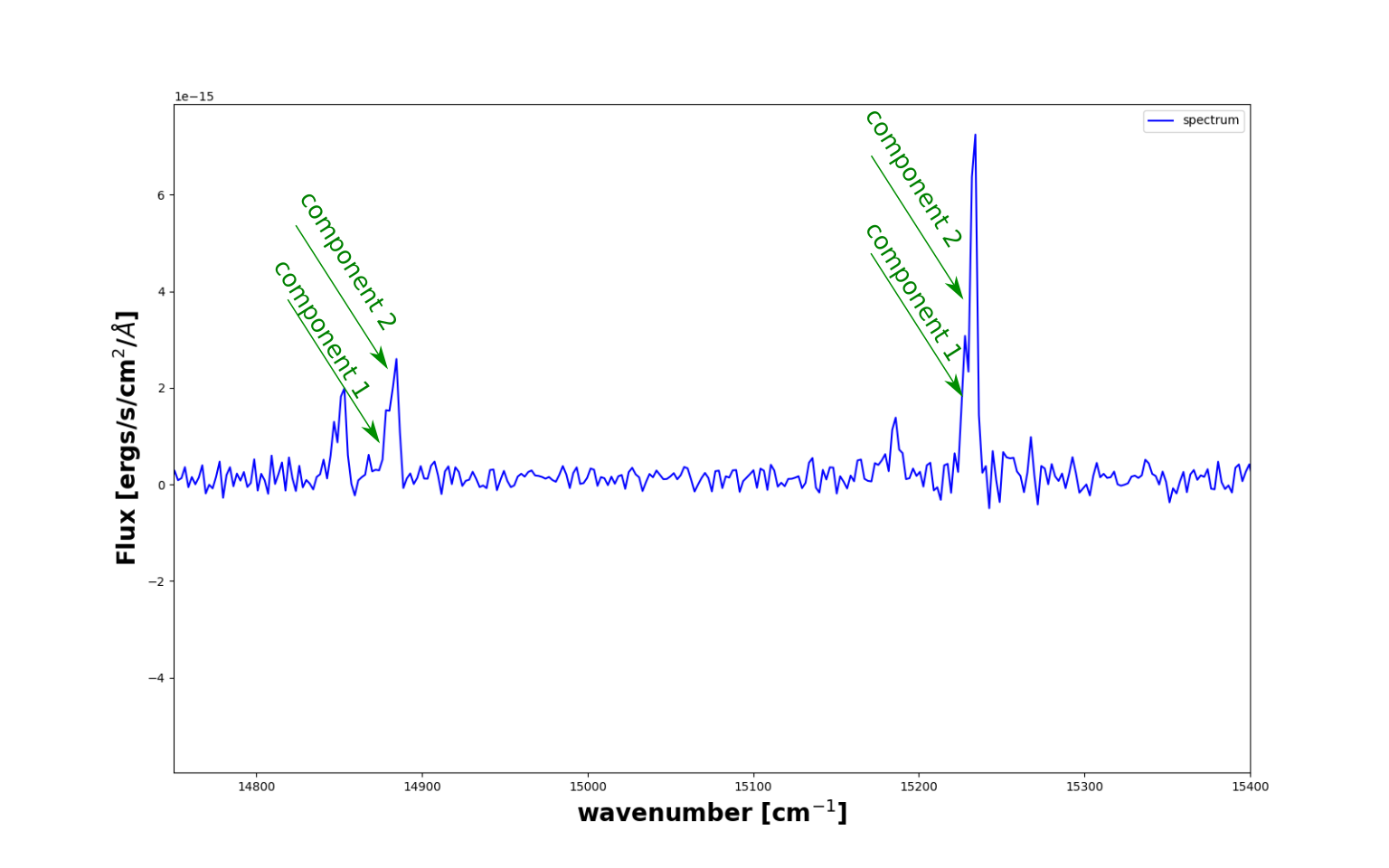}
    \caption{Zoom-in on the SN3 spectrum of region 27. We have noted the location of the double components with green arrows.}
    \label{fig:id27}
\end{figure}

In fitting the two H$\alpha$ components and two \sii{} doublet components, we were able to distinguish emission lines from each component. We note that the flux of the first component of the \sii{} doublet and H$\alpha$ emission line is less than the second component. Moreover, the combined flux of the first \sii{} doublet component is greater than the H$\alpha$ emission line which indicates component 1 represents the SNR emission while component 2 represents the \hii{} region emission. This finding confirms our hypothesis that \hii{} region or DIG contamination is responsible for the misclassifications. 

In artificially reducing the spectral resolution to 3000, we smear the double components into a single component. Although this clearly ignores the underlying astrophysics of the region, it allows for a more succinct fit of the region.
We apply the network assuming these updated values for the SN3 emission lines; however, the network still classifies the region as an \hii{} region. This fortifies our position that contaminant emission from an overlapping \hii{} region or DIG is causing a misclassification of the supernova remnant. 

The careful review of our results can help appreciate the level of contamination of the emission line spectra in galaxies even in the case where spatial resolution enables us to identify the complex structures in the ISM. Moreover, it demonstrates that the decontamination of spectra is not trivial and using dynamical properties such as Full Width Half Maximum (FWHM) and decomposition of the different velocity components of the ionized gas offer a new avenue to identify the well blended SNR emission nebulae.



\section{SNR Misclassifications}

\subsection{Explanation of Misclassifications}\label{app:miss}
In addition to misclassifications of supernova remnants due to \hii{} or DIG contamination, a careful examination of their spectra revealed additional reasons for the incorrect classifications. Two SN2 spectra (IDs 102 and 125) demonstrate a high level of noise not present in the SN3 filter (where our signal-to-noise ratio threshold was applied); the misclassifications are likely due to the poor constraints on \oiii{ and H$\beta$ due to the noise. An additional three SN2 spectra (IDs 58, 75,  and 98) reveal a complete lack of \oiii{}}$\lambda$5007 emission; again, this explains the misclassification by the network. Therefore, if we apply an additional signal-to-noise threshold to SN1 and SN2, we have only 3 misclassified supernova remnants instead of 8. This brings the accuracy from 64\% to 82\%. Similarly, the misclassified planetary nebula would be removed.

\subsection{Interpolating over the SNR grid}\label{app:miss-interp}
\begin{figure}
    \centering
    \includegraphics[width=0.49\textwidth]{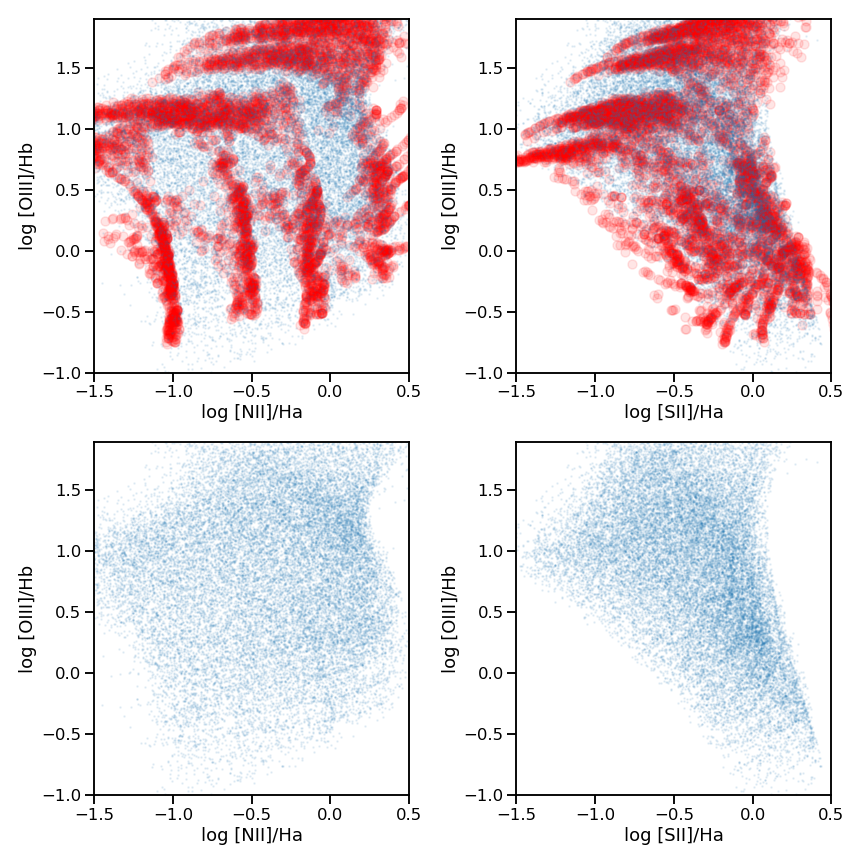}
    \caption{Line-ratio parameter space showing the initial data (in red) and the interpolated data (in blue).}
    \label{fig:interpolation}
\end{figure}

As discussed in section $\S$\ref{sec:M33}, the SNR grid from the 3MdB shock extension has gaps in the line-ratio parameter space. In figure \ref{fig:interpolation}, we show the initial data (in red) and the interpolated data (in blue). The top two figures juxtapose the data initially collected from the 3MdB shock extension in red and the interpolated data in blue in the two standard BPT diagrams. The bottom panels show only the interpolated data points.
The figure demonstrates that our linear interpolation scheme covers the entire parameter space.

\bsp	
\label{lastpage}
\end{document}